\documentclass[12pt]{article}

\usepackage{graphicx}
\usepackage{natbib}
\usepackage{epsfig}
\usepackage{amsfonts} 
\usepackage{amsmath}
\usepackage{amssymb,longtable,enumitem}
\usepackage{mathtools}
\usepackage{mathrsfs}
\usepackage{setspace}
\usepackage{hyperref}
\usepackage{tabularx}
\usepackage{booktabs}
\usepackage{float}
\usepackage[margin=1in]{geometry}
\usepackage{bm}
\usepackage{etoc}

\date{}
\LTcapwidth=\textwidth

\newcommand{\trans}{\top}
\newcommand{\diag}{\operatorname{diag}}
\newcommand{\blockdiag}{\operatorname{blockdiag}}
\newcommand{\tr}{\operatorname{tr}}

\newcommand{\R}{\mathbb{R}}
\newcommand{\one}{\boldsymbol{1}}

\begin{document}
\etocdepthtag.toc{main}
\doublespacing

\title{Latent-Class Poisson-Lognormal Mixed Models for Clustering Multivariate Longitudinal Count Data}

\author{Pengqi Liu\footnote{School of Mathematics and Statistics, Carleton University, 1125 Colonel By Drive, K1S 5B6, Ottawa, Ontario, Canada. e: PengqiLiu@cunet.carleton.ca} \and Utkarsh J. Dang \footnote{Department of Health Sciences, Carleton University, 1125 Colonel By Drive, K1S 5B6, Ottawa, Ontario, Canada. e: UtkarshDang@cunet.carleton.ca} \and Sanjeena Subedi \footnote{School of Mathematics and Statistics, Carleton University, 1125 Colonel By Drive, K1S 5B6, Ottawa, Ontario, Canada. e: SanjeenaDang@cunet.carleton.ca}}

\maketitle

\begin{abstract}
\textbf{Motivation:} Multivariate longitudinal count data arise in biological studies, where responses may be over-dispersed, correlated within subjects and over time, measured at irregular times, and exhibit unobserved subgroup structure. Our goal was to develop a model that can accommodate over-dispersion, longitudinal dependence, cross-outcome dependence, covariate effects, irregular observation times, and covariate-dependent latent-class membership. \\
\textbf{Results:} We propose a latent-class Poisson-lognormal mixed model for clustering heterogeneous multivariate count trajectories. Estimation is performed using a variational EM algorithm designed for scalable inference. Simulation studies show accurate parameter estimation, strong latent-class recovery, and reliable selection of the number of classes, as compared to state-of-the-art approaches. We also applied the proposed approach to longitudinal neutrophil RNA-seq data from hospitalized COVID-19 patients as an illustration. The proposed framework provides a flexible approach for clustering complex longitudinal biological count data.\\
\textbf{Availability and Implementation:} Implementation code is available at this \mbox{\href{https://github.com/PengqiLiu-Mc/latent-class-mpln-mixed-models}{GitHub repository}}.\\
\textbf{Contact:}
\href{mailto:PengqiLiu@cunet.carleton.ca}{PengqiLiu@cunet.carleton.ca};
\href{mailto:SanjeenaDang@cunet.carleton.ca}{SanjeenaDang@cunet.carleton.ca}\\
\textbf{Supplementary information:}
Supplementary material accompanies this manuscript.
\end{abstract}

\textbf{Keywords}: mixture models, MPLN distribution, longitudinal data, discrete count data, RNA-seq data.

\section{Introduction}
\label{sec: Intro}
Longitudinal multivariate count data are increasingly common in modern biological and health studies. In longitudinal RNA sequencing (RNA-seq), immune profiling, microbiome studies, and related applications, multiple count-valued markers may be measured repeatedly for each case/subject over time. Such data provide an opportunity to study dynamic biological processes, such as disease progression, treatment response, and immune regulation, while identifying latent subgroups of subjects with distinct temporal patterns. However, analysis of such data comes with several statistical challenges. Biological count responses are discrete and typically exhibit variability (over-dispersion) that exceeds what is expected under a Poisson model \citep{subedi2025multivariate}. Other important aspects include repeated measurements from the same subject being serially correlated, markers measured at the same visit possibly cross-correlated, subjects frequently observed at irregular visit times (i.e., at subject-specific time points rather than on a common schedule), and records may be incomplete. These features motivate the development of a model-based method for clustering the observed counts directly while accommodating the longitudinal sampling structure without relying on transformations, alignment to a common time grid, or pre-imputation of missing visit data.

The negative binomial (NB) distribution has become a standard choice for over-dispersed biological count data, particularly for RNA-seq analysis via {\tt R} packages such as {\sf edgeR} and {\sf DESeq2} \citep{robinson2010edger, love2014moderated}. These methods have been influential for gene-wise differential expression and count regression. However, standard NB RNA-seq workflows are typically gene-wise or marker-wise: they model over-dispersion for each response but do not directly model the joint dependence structure among multiple outcomes. Furthermore, although the NB distribution provides an effective model for univariate over-dispersed counts, extending Poisson-based count models to the multivariate setting often introduces substantial computational difficulty or restrictive dependence structures. For example, some multivariate Poisson formulations allow only non-negative dependence \citep{inouye2017review}, while independence assumptions are often adopted in multivariate count analyses for computational convenience \citep{si2014model, rau2015co, papastamoulis2016estimation}. Such simplifications are problematic for modern biological data, where correlations among transcripts, taxa, or biomarkers may be scientifically meaningful and may include both positive and negative associations.

The multivariate Poisson-lognormal (MPLN) distribution provides a natural foundation for modeling dependent over-dispersed count data. In the MPLN model, conditionally independent Poisson variables are linked through latent Gaussian variables on the log-mean scale \citep{aitchison1989multivariate}. This hierarchical construction preserves the count scale, accommodates zero counts and over-dispersion, and permits flexible signed dependence among outcomes. Recent work has demonstrated the usefulness of MPLN models for biological count data, including multivariate RNA-seq clustering \citep{silva2019multivariate}, parsimonious MPLN mixture models using covariance constraints \citep{subedi2020family}, matrix-variate MPLN mixtures for three-way RNA-seq read count data \citep{silva2023finite}, Poisson-lognormal factor-analytic mixtures for high-dimensional count clustering \citep{payne2025finite}, and flexible Poisson-lognormal frameworks for joint multivariate count analysis \citep{chiquet2021poisson}. 

Nevertheless, existing MPLN clustering methods do not fully address flexible longitudinal data structures. Most MPLN mixture models are formulated for multivariate count vectors, such as genes or samples measured under a fixed set of conditions, and are primarily used for clustering expression profiles, microbiome samples, or high-dimensional count vectors \citep{silva2019multivariate}. They do not provide a subject-level longitudinal regression framework in which each subject may have their own measurement schedule, time-varying covariates, repeated multivariate outcomes, and latent-class membership. On the other hand, existing Gaussian longitudinal clustering methods and latent-class mixed models provide powerful tools for heterogeneous trajectories, including parsimonious covariance modeling and latent-class structure \citep{mcnicholas2010model, proust2017estimation}. However, they are not directly tailored to over-dispersed multivariate count responses. In addition, many longitudinal clustering methods are most naturally formulated for measurements collected on a common regular grid \citep{genolini2010kml, mcnicholas2010model}, so irregular visits and incomplete records often require alignment, interpolation, or imputation. Recent multivariate count generalized linear mixed models can model correlated count outcomes with covariate effects using Poisson, NB, or COM-Poisson distributions \citep{daSilva2024MGLMM}, but they are not primarily designed for latent-class clustering of subjects with class-specific longitudinal trajectories, serial dependence, and irregular observation schedules.

In this paper, we propose a latent-class Poisson-lognormal mixed model for heterogeneous multivariate longitudinal count data with subject-specific irregular visit times. Our proposed framework clusters subjects according to distinct multivariate count trajectories while accommodating over-dispersion and dependence within and across outcomes. Estimation is performed using a variational EM algorithm. Hence, the main contributions of this work are threefold. First, we introduce a latent-class Poisson-lognormal mixed regression framework for subject-level clustering of multivariate longitudinal count data, allowing subjects to be observed at subject-specific visit times with possibly incomplete records. Second, we develop an interpretable model structure that jointly captures over-dispersion, class-specific covariate effects, serial dependence, and cross-outcome dependence, while allowing covariates to inform latent-class membership. Third, we develop a variational EM algorithm for fast, scalable parameter estimation and latent-class inference. Through simulation studies and a real-data application to longitudinal neutrophil RNA-seq count data from hospitalized COVID-19 patients \citep{lasalle2022longitudinal}, we demonstrate that the proposed method can recover heterogeneous multivariate count trajectories and provides an interpretable tool for studying biological heterogeneity in longitudinal count data.

\section{Model and methods}
\label{sec:methods}

The multivariate Poisson-lognormal (MPLN) model assigns a multivariate Gaussian distribution to the log-intensities of conditionally independent Poisson variables \citep{aitchison1989multivariate}. Integrating over the Gaussian layer accommodates extra-Poisson variation and induces a rich cross-outcome dependence structure, including both positive and negative associations. Hence, the model handles over-dispersion and dependence while retaining the discrete scale of the observations and allowing covariates and known offsets to enter through the log-abundance. We adapt the MPLN framework to multivariate longitudinal count data by introducing subject-specific random effects and visit-level multivariate Gaussian errors. The random effects capture serial dependence across measurements taken at possibly irregular times, whereas the visit-level errors capture dependence among outcomes measured concurrently. The resulting Poisson-lognormal mixed regression model is further extended to a latent-class structure in Section~\ref{subsec:latent_class_model}.

\subsection{Single-class Poisson-lognormal mixed regression model}
\label{subsec:plnmm}

We first describe the single-class Poisson-lognormal mixed regression model.
Consider $n$ subjects and $p$ count outcomes. Subject $i$ is observed at
subject-specific visit times $t_{i1}<\cdots<t_{iT_i}$, where the number and
timing of visits may vary across subjects. Let
$\boldsymbol{Y}_{il}=(Y_{i1l},\ldots,Y_{ipl})^\top$ denote the multivariate
count response at visit time $t_{il}$. For outcome $j$, let
$\boldsymbol{X}_{ij}(t_{il})$ and $\boldsymbol{R}_{ij}(t_{il})$ be,
respectively, the fixed- and random-effect design vectors evaluated at the
observed visit time. These designs may contain functions of time and
time-varying covariates. 

For $i=1,\ldots,n$, $j=1,\ldots,p$, and $l=1,\ldots,T_i$,
\begin{align}
 Y_{ijl}\mid \eta_{ijl}
 &\sim \operatorname{Poisson}\{\exp(\eta_{ijl})\}, \label{eq:poisson_layer}\\
 \eta_{ijl}
 &=o_{il}+\boldsymbol{X}_{ij}(t_{il})^\top\boldsymbol{\beta}_{j}
   +\boldsymbol{R}_{ij}(t_{il})^\top\boldsymbol{b}_{ij}
   +\epsilon_{ijl}. \label{eq:linear_predictor_single}
\end{align}
Here $o_{il}$ is an optional known offset, commonly used in RNA-seq applications to account for sample-specific sequencing depth; it is set to zero when no offset is required. One convention sets $o_{il}=\log(s_{il})$, where $s_{il}$ is a relative size factor, as in DESeq2 \citep{love2014moderated}; \cite{silva2019multivariate} used the same size-factor form in an MPLN mixture model, with the factors obtained using edgeR normalization \citep{robinson2010scaling}. In our RNA-seq application, we set $o_{il}=\log(L_{il})$, where $L_{il}$ is the edgeR effective library size, defined as the raw library size multiplied by its normalization factor \citep{ChenEtAl2025}. These formulations follow the same underlying offset principle: both represent sequencing depth as a known multiplicative exposure on the count scale. If all samples have equal sequencing depth and require no additional composition adjustment, then $L_{il} \equiv L$ and $s_{il} \equiv 1$, so the common offset is either absorbed into the fixed-effect intercept or equals zero, and no explicit sequencing-depth adjustment is needed.
Throughout, $\eta_{ijl}$ denotes the overall log-abundance, including the offset. Because $o_{il}$ is fixed and known, it shifts the conditional mean
but does not alter the covariance structure or introduce an additional model parameter.

The outcome-specific random effects are stacked as
\begin{align*}
 \boldsymbol{b}_{i}
 &=(\boldsymbol{b}_{i1}^\top,\ldots,\boldsymbol{b}_{ip}^\top)^\top
 \sim N(\boldsymbol{0},\boldsymbol{\Omega}),\\
 \boldsymbol{\Omega}
 &=\operatorname{blockdiag}\{\boldsymbol{\Omega}^{(1)},\ldots,
 \boldsymbol{\Omega}^{(p)}\}.
\end{align*}
At each visit,
$\boldsymbol{\epsilon}_{il}
=(\epsilon_{i1l},\ldots,\epsilon_{ipl})^\top
\sim N(\boldsymbol{0},\boldsymbol{\Sigma})$, where
$\boldsymbol{\Sigma}$ is unrestricted and positive definite. We assume that
$\boldsymbol{\epsilon}_{il}$ is independent across visits and independent of
$\boldsymbol{b}_i$.
This construction separates two scientifically distinct sources of
dependence. The random effects induce serial association for repeated
measurements of the same outcome, whereas $\boldsymbol{\Sigma}$ captures
cross-outcome association among markers measured at the same visit. In
particular,
\begin{align}
 &\operatorname{Cov}(\eta_{ijl},\eta_{ij'l'})
 \nonumber\\
 &\quad =
 \mathbb{I}(j=j')\,
 \boldsymbol{R}_{ij}(t_{il})^\top
 \boldsymbol{\Omega}^{(j)}
 \boldsymbol{R}_{ij'}(t_{il'})
 +\mathbb{I}(l=l')\,\Sigma_{jj'} . \label{eq:latent_covariance}
\end{align}
The Gaussian variation on the log-mean scale induces extra-Poisson variation, allowing the model to be fitted directly to over-dispersed counts.

Note that with this construction, measurements at the actual $t_{il}$ are used, and neither a common time grid nor interpolation are required. We assume that all $p$ outcomes are observed at each recorded visit, but not all participants are measured regularly at the same time points. We condition on the realized visit schedules and assume that the observation
process is non-informative given the recorded covariates.

\subsection{Latent-class Poisson-lognormal mixed regression model}
\label{subsec:latent_class_model}

To represent unobserved population heterogeneity, let
$Z_i\in\{1,\ldots,G\}$ denote the latent class of subject $i$. Conditional on
$Z_i=g$, we replace the parameters in
\eqref{eq:linear_predictor_single} by class-specific quantities:
\begin{align}
 &Y_{ijl}\mid Z_i=g,\eta_{ijlg}
 \sim \operatorname{Poisson}\{\exp(\eta_{ijlg})\}, \nonumber\\
 &\eta_{ijlg}
 =o_{il}+\boldsymbol{X}_{ij}(t_{il})^\top\boldsymbol{\beta}_{gj}
+\boldsymbol{R}_{ij}(t_{il})^\top\boldsymbol{b}_{ijg}
   +\epsilon_{ijlg}, \label{eq:mixture_model}
\end{align}
where
\begin{align*}
 \boldsymbol{b}_{ig}
 &=(\boldsymbol{b}_{i1g}^\top,\ldots,\boldsymbol{b}_{ipg}^\top)^\top
 \sim N(\boldsymbol{0},\boldsymbol{\Omega}_g),\\
 \boldsymbol{\Omega}_g
 &=\operatorname{blockdiag}\{\boldsymbol{\Omega}^{(1)}_g,\ldots,
 \boldsymbol{\Omega}^{(p)}_g\},
\end{align*}
and
$\boldsymbol{\epsilon}_{ilg}
=(\epsilon_{i1lg},\ldots,\epsilon_{iplg})^\top
\sim N(\boldsymbol{0},\boldsymbol{\Sigma}_g)$. The class-specific
$\boldsymbol{\beta}_{gj}$, $\boldsymbol{\Omega}_g$, and
$\boldsymbol{\Sigma}_g$ allow latent subgroups to differ in their
multivariate trajectories, serial variation, and cross-outcome association.

Class membership may depend on a vector of time-independent subject-level
covariates $\boldsymbol{c}_i$. Let
$\widetilde{\boldsymbol{c}}_i=(1,\boldsymbol{c}_i^\top)^\top$. Using class
$G$ as the reference, a Softmax gating model specifies, for $g = 1,\ldots,G-1$,
\begin{align}
 \pi_{ig}
 &=\Pr(Z_i=g\mid\boldsymbol{c}_i)
 =\frac{\exp(\widetilde{\boldsymbol{c}}_i^\top\boldsymbol{\xi}_g)}
 {1+\sum_{h=1}^{G-1}
 \exp(\widetilde{\boldsymbol{c}}_i^\top\boldsymbol{\xi}_h)}, \nonumber\\
 \text{and } \pi_{iG}
 &=\frac{1}
 {1+\sum_{h=1}^{G-1}
 \exp(\widetilde{\boldsymbol{c}}_i^\top\boldsymbol{\xi}_h)} .
 \label{eq:softmax_gate}
\end{align}
An intercept-only gate gives constant mixing proportions as a special case.
In the observed design information $\mathcal{D}_i$ where
\begin{equation*}
    \mathcal{D}_i = \{o_{il}, t_{il}, \boldsymbol{X}_{ij}(t_{il}), \boldsymbol{R}_{ij}(t_{il}): l = 1,\ldots,T_i, j = 1,\ldots,p\},
\end{equation*}
all elements are treated as given. The subject-level likelihood is therefore a finite mixture
\begin{equation}
    p(\boldsymbol{Y}_i \mid \boldsymbol{c}_i, \mathcal{D}_i) = \sum_{g=1}^{G} \pi_{ig}(\boldsymbol{c}_i) p_g(\boldsymbol{Y}_i \mid \mathcal{D}_i),
    \label{eq:obs_likelihood}
\end{equation}
where 
$
    \boldsymbol{Y}_i = \left(\boldsymbol{Y}_{i1}^\top,\ldots, \boldsymbol{Y}_{iT_i}^\top\right)^\top$
contains all outcomes recorded at the subject's observed visits. The class-specific marginal distribution is
\begin{align}
    p_g(\boldsymbol{Y}_i \mid \mathcal{D}_i) &= \iint \prod_{l=1}^{T_i} \prod_{j=1}^{p} f_{\text{Pois}} \left(Y_{ijl}; \exp \{\eta_{ijlg}\} \right) \nonumber\\
    &\hspace{24pt} \times \phi(\boldsymbol{b}_{ig}; \boldsymbol{0}, \boldsymbol{\Omega}_g) \prod_{l=1}^{T_i} \phi(\boldsymbol{\epsilon}_{ilg}; \boldsymbol{0}, \boldsymbol{\Sigma}_g) d \boldsymbol{\epsilon}_{ig} d \boldsymbol{b}_{ig},
    \label{eq:marginal_likelihood}
\end{align}
where $\boldsymbol{\epsilon}_{ig} = (\boldsymbol{\epsilon}_{i1g}^\top, \ldots, \boldsymbol{\epsilon}_{iT_ig}^\top)^\top$, $f_{\text{Pois}}(y; \lambda)$ denotes the Poisson probability mass function with mean $\lambda$, and $\phi(\cdot; \boldsymbol{\mu}, \boldsymbol{V})$ denotes a (multivariate) Gaussian density with mean $\boldsymbol{\mu}$ and covariance matrix $\boldsymbol{V}$.
Hence, the model deals with raw count modeling, extra-Poisson variation, serial and cross-outcome dependence, irregular observation times, incomplete records, time-varying covariates, and latent trajectory heterogeneity. 

\subsection{Variational estimation and latent-class inference}
\label{subsec:vem}

Estimation is challenging for three main reasons. First, each
class-specific marginal likelihood \eqref{eq:marginal_likelihood} requires integration over the
subject-specific random effects and visit-level Gaussian latent variables,
with a dimension that increases with $p$ and $T_i$. Second, the Poisson likelihood is
non-conjugate with the Gaussian latent distribution, so neither the
component likelihood nor the latent posterior are available in closed form.
Third, the unknown class indicator $Z_i$ adds a discrete latent layer, coupling parameter estimation and class-membership inference.

We address these challenges using a variational EM (VEM) algorithm
with variational Gaussian approximations (VGA)
\citep{Bishop2006, WainwrightJordan2008, ArridgeEtAl2018}. Related VGA-based VEM procedures have been successfully applied to mixtures of MPLN distributions, including parsimonious covariance and factor analyzers \citep{subedi2020family, payne2025finite}. All distributions below are conditional on the observed design information $\mathcal{D}_i$ defined above, and suppressed for concision. Let
$\boldsymbol{\eta}_{ilg}
=(\eta_{i1lg},\ldots,\eta_{iplg})^\top$ denote the vector of latent log-intensities at visit $l$, for $l=1,\ldots,T_i$, and define
$\boldsymbol{\eta}_{ig}
=(\boldsymbol{\eta}_{i1g}^{\top},\ldots,
\boldsymbol{\eta}_{iT_i g}^{\top})^{\top}$ as the corresponding subject-level stacked vector. Let
$\boldsymbol{X}_i$ and $\boldsymbol{M}_i$ denote the corresponding stacked fixed- and random-effect design matrices. Conditional on
$\boldsymbol{b}_{ig}$ and $Z_i=g$,
\begin{align}
 \boldsymbol{\eta}_{ig}\mid\boldsymbol{b}_{ig},Z_i=g
 &\sim N\{\boldsymbol{\nu}_{ig}(\boldsymbol{b}_{ig}),
          \boldsymbol{K}_{\Sigma,ig}\}, \nonumber\\
 \boldsymbol{\nu}_{ig}(\boldsymbol{b}_{ig})
 &=\boldsymbol{X}_i\boldsymbol{\beta}_g
   +\boldsymbol{M}_i\boldsymbol{b}_{ig}
   +\boldsymbol{o}_i, \nonumber\\
 \boldsymbol{K}_{\Sigma,ig}
 &=\boldsymbol{I}_{T_i}\otimes\boldsymbol{\Sigma}_g,
 \label{eq:stacked_latent_model}
\end{align}
where
$\boldsymbol{o}_i=(o_{i1}\boldsymbol{1}_p^\top,\ldots,
o_{iT_i}\boldsymbol{1}_p^\top)^\top$ is the known stacked offset.

Let $\tau_{ig}=q_i(Z_i=g)$. Given $Z_i=g$, we approximate the posterior of
$(\boldsymbol{b}_{ig},\boldsymbol{\eta}_{ig})$ by
\begin{align}
 q_{ig}(\boldsymbol{b}_{ig},\boldsymbol{\eta}_{ig})
 &=q_{ig}(\boldsymbol{b}_{ig})
   q_{ig}(\boldsymbol{\eta}_{ig}\mid\boldsymbol{b}_{ig}), \nonumber\\
 q_{ig}(\boldsymbol{b}_{ig})
 &=N(\boldsymbol{\mu}^{(b)}_{ig},
      \boldsymbol{S}^{(b)}_{ig}), \nonumber\\
 q_{ig}(\boldsymbol{\eta}_{ig}\mid\boldsymbol{b}_{ig})
 &=N(\boldsymbol{A}_{ig}
      +\boldsymbol{L}_{ig}\boldsymbol{b}_{ig},
      \boldsymbol{V}_{ig}).
 \label{eq:flexible_variational_family}
\end{align}
Both $\boldsymbol{S}^{(b)}_{ig}$ and $\boldsymbol{V}_{ig}$ may be unrestricted, and
the free matrix $\boldsymbol{L}_{ig}$ captures posterior dependence between
the random effects and latent log-intensities. This flexibility concerns the
variational posterior only: $\boldsymbol{\Omega}_g$ remains block-diagonal
across outcomes, and the model covariance in
\eqref{eq:stacked_latent_model} remains
$\boldsymbol{I}_{T_i}\otimes\boldsymbol{\Sigma}_g$.

For subject $i$ and class $g$, define
\begin{align}
 B_{ig}
 &=
 E_{q_{ig}}\!\left[
 \log p(\boldsymbol{Y}_i,\boldsymbol{b}_{ig},
        \boldsymbol{\eta}_{ig}\mid Z_i=g;\boldsymbol{\theta}_g)
 \right] \nonumber\\
 &\quad{}
 -E_{q_{ig}}\!\left[
 \log q_{ig}(\boldsymbol{b}_{ig},\boldsymbol{\eta}_{ig})
 \right] \nonumber\\
 &\leq \log p_g(\boldsymbol{Y}_i\mid\mathcal{D}_i),
 \label{eq:component_elbo}
\end{align}
where
$\boldsymbol{\theta}_g
=(\boldsymbol{\beta}_g,\boldsymbol{\Omega}_g,
\boldsymbol{\Sigma}_g)$. The mixture evidence lower bound (ELBO) is
\begin{equation}
 \mathcal{L}(q,\boldsymbol{\theta})
 =
 \sum_{i=1}^{n}\sum_{g=1}^{G}
 \tau_{ig}
 \{\log\pi_{ig}-\log\tau_{ig}+B_{ig}\},
 \label{eq:mixture_elbo}
\end{equation}
where
$\boldsymbol{\theta}
=\{\boldsymbol{\beta}_g,\boldsymbol{\Omega}_g,
\boldsymbol{\Sigma}_g\}_{g=1}^{G}
\cup\{\boldsymbol{\xi}_g\}_{g=1}^{G-1}$.
For fixed $\boldsymbol{\theta}$, maximizing
\eqref{eq:mixture_elbo} over $q$ is equivalent to minimizing its
Kullback--Leibler divergence from the true latent posterior.

Each VEM iteration comprises three updates:

\underline{\emph{V-step}}: For every
$(i,g)$, we update the joint Gaussian approximation to
$\boldsymbol{U}_{ig}
=(\boldsymbol{b}_{ig}^{\top},\boldsymbol{\eta}_{ig}^{\top})^{\top}$.
The model-implied Gaussian prior mean of $\boldsymbol{U}_{ig}$ is
$\{\boldsymbol{0}^{\top},
(\boldsymbol{X}_i\boldsymbol{\beta}_g+\boldsymbol{o}_i)^{\top}\}^{\top}$. The joint mean is
updated by Newton's method, with step halving when needed to avoid a decrease
in $B_{ig}$, and the joint covariance is updated by a fixed-point iteration.
The resulting moments are converted to the conditional parameterization in
\eqref{eq:flexible_variational_family} using Gaussian conditional identities.
Details are provided in the Supplementary Material.

\underline{\emph{E-step}}: The approximate posterior class probabilities are
updated by
\begin{equation}
 \tau_{ig}
 =
 \frac{\pi_{ig}\exp(B_{ig})}
 {\sum_{h=1}^{G}\pi_{ih}\exp(B_{ih})},
 \qquad g=1,\ldots,G.
 \label{eq:responsibility_update}
\end{equation} 
Thus, $\tau_{ig}$ combines the
prior class probability from the gating model with the variational evidence
for the subject's observed trajectory.

\underline{\emph{M-step}}: The population parameters are updated using weighted variational moments. Define the offset-adjusted
pseudo-response
\[
 \overline{\boldsymbol{\eta}}_{ig}^{\,*}
 =
 \boldsymbol{A}_{ig}-\boldsymbol{o}_i
 +(\boldsymbol{L}_{ig}-\boldsymbol{M}_i)
  \boldsymbol{\mu}^{(b)}_{ig}.
\]
Using the current $\boldsymbol{K}_{\Sigma,ig}$, the fixed effects are updated
by weighted generalized least squares:
\begin{align}
 \boldsymbol{\beta}_g^{\mathrm{new}}
 &=
 \left\{\sum_{i=1}^{n}\tau_{ig}
 \boldsymbol{X}_i^\top\boldsymbol{K}_{\Sigma,ig}^{-1}
 \boldsymbol{X}_i\right\}^{-1}
 \sum_{i=1}^{n}\tau_{ig}
 \boldsymbol{X}_i^\top\boldsymbol{K}_{\Sigma,ig}^{-1}
 \overline{\boldsymbol{\eta}}_{ig}^{\,*}.
 \label{eq:beta_update}
\end{align}
Let $N_g=\sum_i\tau_{ig}$, and let
$\boldsymbol{\mu}^{(b)}_{ijg}$ and
$\boldsymbol{S}^{(b)}_{ig,jj}$ be the outcome-$j$ subvector and diagonal
outcome block of the variational random-effect moments. The block-diagonal
random-effect covariance is updated as
\begin{equation}
 \boldsymbol{\Omega}_g^{(j),\mathrm{new}}
 =
 \frac{1}{N_g}\sum_{i=1}^{n}\tau_{ig}
 \left\{\boldsymbol{S}^{(b)}_{ig,jj}
 +\boldsymbol{\mu}^{(b)}_{ijg}
  \boldsymbol{\mu}^{(b)\top}_{ijg}\right\}.
 \label{eq:omega_update}
\end{equation}
After updating $\boldsymbol{\beta}_g$, define
\begin{align*}
 \boldsymbol{d}_{ig}
 &=
 \boldsymbol{A}_{ig}-\boldsymbol{o}_i
 -\boldsymbol{X}_i\boldsymbol{\beta}_g^{\mathrm{new}}
 +(\boldsymbol{L}_{ig}-\boldsymbol{M}_i)
  \boldsymbol{\mu}^{(b)}_{ig},\\
 \boldsymbol{\Psi}_{ig}
 &=
 \boldsymbol{V}_{ig}
 +(\boldsymbol{L}_{ig}-\boldsymbol{M}_i)
  \boldsymbol{S}^{(b)}_{ig}
  (\boldsymbol{L}_{ig}-\boldsymbol{M}_i)^\top
 +\boldsymbol{d}_{ig}\boldsymbol{d}_{ig}^{\top}.
\end{align*}
Partition $\boldsymbol{\Psi}_{ig}$ into $p\times p$ visit blocks and denote
its $l$th diagonal block by $\boldsymbol{\Psi}_{ig,ll}$. Then
\begin{equation}
 \boldsymbol{\Sigma}_g^{\mathrm{new}}
 =
 \frac{\sum_{i=1}^{n}\tau_{ig}
       \sum_{l=1}^{T_i}\boldsymbol{\Psi}_{ig,ll}}
      {\sum_{i=1}^{n}\tau_{ig}T_i}.
 \label{eq:sigma_update}
\end{equation}
Only diagonal visit blocks enter \eqref{eq:sigma_update}, consistently with
the assumed independence of the visit-level Gaussian errors across time.
Finally, the gating coefficients maximize
$\sum_i\sum_g\tau_{ig}\log\pi_{ig}$, equivalent to multinomial logistic
regression with soft class labels. For an intercept-only gate, $\pi_g^{\mathrm{new}}=N_g/n$.

The V-, E-, and M-steps are repeated until the absolute change in the ELBO between two successive iterations falls below $10^{-4}$. At convergence, the population-parameter estimates address the first inferential target, while
$
 \widehat{\tau}_{ig}
 \approx
 \Pr_{\widehat{\boldsymbol{\theta}}}
 (Z_i=g\mid\boldsymbol{Y}_i,\boldsymbol{c}_i,\mathcal{D}_i)
$
quantifies uncertainty in latent-class membership. When a hard partition is
required, we use the variational maximum \textit{a posteriori} assignment
$
 \widehat{Z}_i
 =
 \arg\max_{g=1,\ldots,G}\widehat{\tau}_{ig}.
$
To reduce sensitivity to local optima,
we use multiple data-adaptive starts and retain the converged solution with the largest ELBO. Details about initialization, numerical stabilization, and estimation procedure are in the Supplementary Material.

\subsection{Model selection and performance assessment}
\label{subsec:model_selection}

For a candidate model specification $\mathcal{M}$, the classical Bayesian information criterion (BIC) is
\begin{equation}
 \operatorname{BIC}(\mathcal{M})
 =
 -2\ell_{\mathcal{M}}
 \bigl(\widehat{\boldsymbol{\theta}}_{\mathcal{M}}^{\,\mathrm{ML}}\bigr)
 +\nu_{\mathcal{M}}\log n,
 \label{eq:classical_bic}
\end{equation}
where
$
 \ell_{\mathcal{M}}(\boldsymbol{\theta}_{\mathcal{M}})
 =
 \sum_{i=1}^{n}
 \log p
 (\boldsymbol{Y}_i\mid\boldsymbol{c}_i,\mathcal{D}_i;
  \boldsymbol{\theta}_{\mathcal{M}})
$
is the observed-data log-likelihood,
$\widehat{\boldsymbol{\theta}}_{\mathcal{M}}^{\,\mathrm{ML}}$ is its maximum-likelihood estimator, and $\nu_{\mathcal{M}}$ is the number of free model parameters \citep{Schwarz1978}. The model with the smallest BIC is preferred.

As the observed-data log-likelihood for the proposed model is intractable, we replace the maximized log-likelihood in \eqref{eq:classical_bic} by the converged mixture
ELBO \citep{subedi2020family, subedi2025multivariate} defined in \eqref{eq:mixture_elbo}. 
Specifically, letting $(\widehat q_{\mathcal{M}},
\widehat{\boldsymbol{\theta}}_{\mathcal{M}})$ denote the converged solution, we use a variational approximation to BIC (variational BIC) defined as
$
 \widetilde{\operatorname{BIC}}(\mathcal{M})
 =
 -2\mathcal{L}_{\mathcal{M}}
 \bigl(\widehat q_{\mathcal{M}},
       \widehat{\boldsymbol{\theta}}_{\mathcal{M}}\bigr)
 +\nu_{\mathcal{M}}\log n.
$
Because the ELBO lower-bounds the observed-data log-likelihood, this is an ELBO-based approximation. We use it to compare candidate models, including the number of latent classes $G$, and select the model with the smallest $\widetilde{\operatorname{BIC}}$. Following the convention for latent-class mixed models implemented in the \texttt{lcmm} package \citep{proust2017estimation}, $n$ denotes the number of independent sampling units (subjects), rather than the total number of longitudinal observations.

In simulations with known labels, latent-class recovery is assessed using the
adjusted Rand index (ARI) \citep{HubertArabie1985}: one indicates identical partitions and values near zero indicate chance-level agreement.

\section{Simulations}
\label{sec:simulations}

\subsection{Estimation, latent-class recovery, and model selection}
\label{subsec:simulations1}

We first evaluated the proposed VEM estimator under data generated from the proposed model. Three simulation settings were considered, each with $100$ independent replications. The single-class setting used $G=1$, $p=3$, and $n\in\{100,300,500\}$ to assess estimation of model parameters $\{\boldsymbol{\beta}, \boldsymbol{\Omega}, \boldsymbol{\Sigma}\}$. Our main investigation used $G=2$, $p=3$, and $n\in\{300,500,700\}$ to assess both parameter estimation and latent-class recovery. Finally, a more challenging $G=3$, $p=2$, $n=500$ setting was used to examine latent-class recovery under weak trajectory separation. 

Subjects were observed at irregular visit times, with the number of possible visits varying across subjects and simulation settings: $T_i \in \{4,5,6\}$ for $G=1$, $T_i \in \{3,4,5,6,7\}$ for $G=2$, and $T_i \in \{2,3,4,5\}$ for $G=3$. For the $G=1$ and $G=2$ settings, the fixed-effect design included an intercept, time, and a time-varying covariate, while the random-effect design included a random intercept and a random time slope. The $G=2$ gating network also included two time-independent covariates. For the $G=3$ setting, the fixed-effect design also included a time-invariant covariate, additionally used in the Softmax gating network. For each replication, candidate models with $G>2$ were fitted from $20$ independently seeded random initializations, and the converged fit attaining the largest final ELBO was retained to reduce sensitivity to local optima. The data-generating mechanism and parameter settings are provided in Supplementary Material.

Figure~\ref{fig:sim_trajectory} shows representative simulated trajectories and illustrates the clustering difficulty using one replication from the $G=2, n=300$ setting and one from the $G=3, n=500$ setting. While the outcomes exhibit distinct class-specific temporal patterns, substantial within-class variation and between-class overlap remain. The $G=3$ setting is especially challenging: the class trajectories overlap considerably for the first outcome and are only moderately separated for the second.

\begin{figure}[t]
\centering

\begin{minipage}{\columnwidth}
\textbf{A}\par\vspace{0.1em}
\centering
\includegraphics[
    width=\linewidth
]{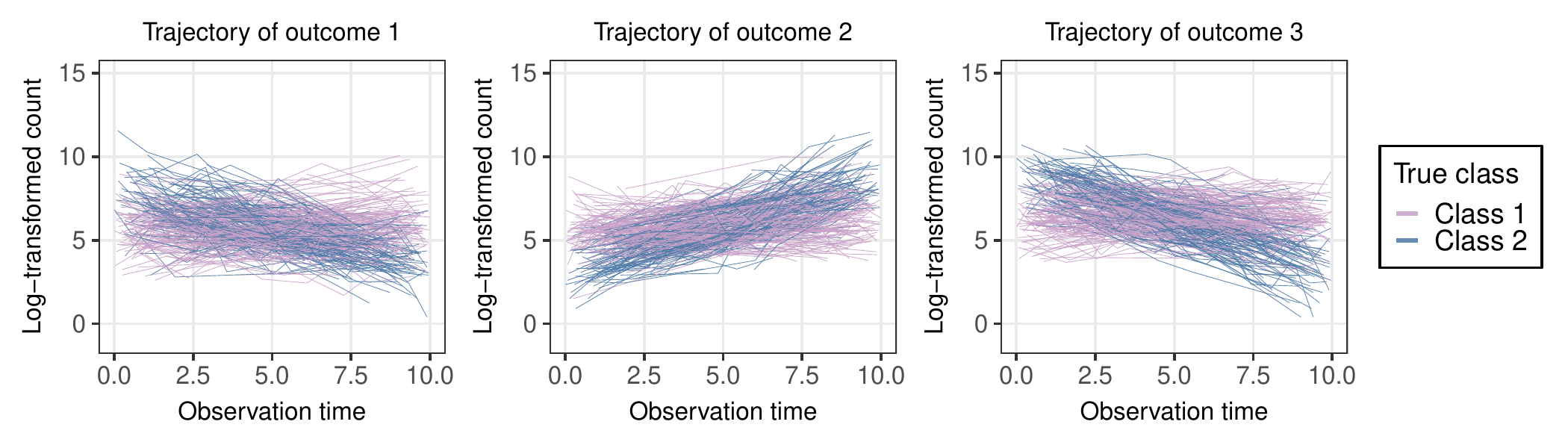}
\end{minipage}

\vspace{0.4em}

\begin{minipage}{\columnwidth}
\textbf{B}\par\vspace{0.1em}
\centering
\includegraphics[
    width=\linewidth
]{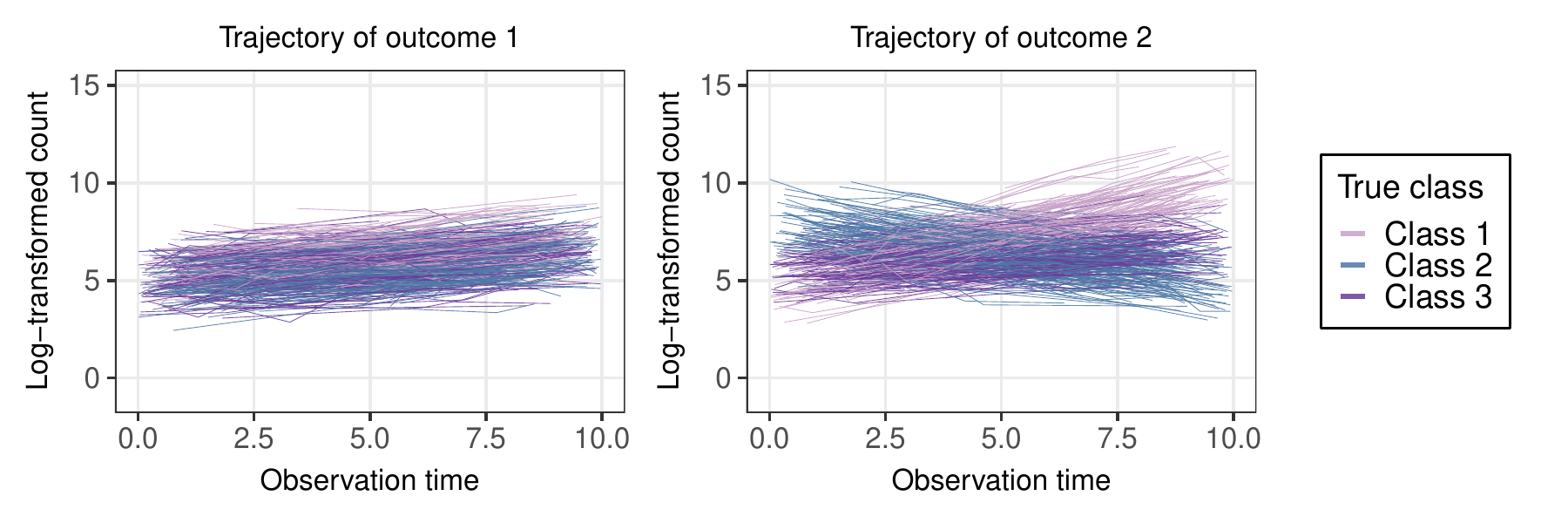}
\end{minipage}

\caption{
\textbf{Representative simulated longitudinal count trajectories.}
\textbf{(A)} Two-class setting with $G=2$, $p=3$, and $n=300$.
\textbf{(B)} Three-class setting with $G=3$, $p=2$, and $n=500$.
The lines represent individual subject trajectories on the log-transformed count scale, and colors indicate the true latent classes.
}
\label{fig:sim_trajectory}
\end{figure}
 
\begin{figure}[H]
\centering

\begin{minipage}{\columnwidth}
\textbf{A}\par\vspace{0.1em}
\centering
\includegraphics[
    width=\linewidth
]{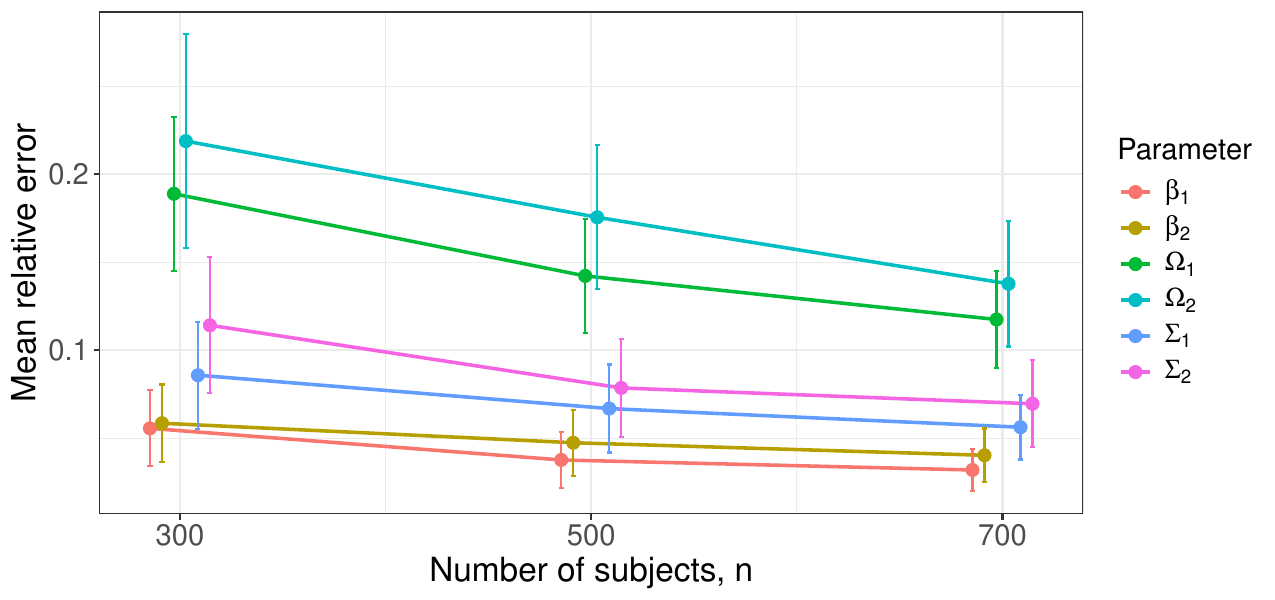}
\end{minipage}

\vspace{0.4em}

\begin{minipage}{\columnwidth}
\textbf{B}\par\vspace{0.1em}
\centering
\includegraphics[
    width=\linewidth
]{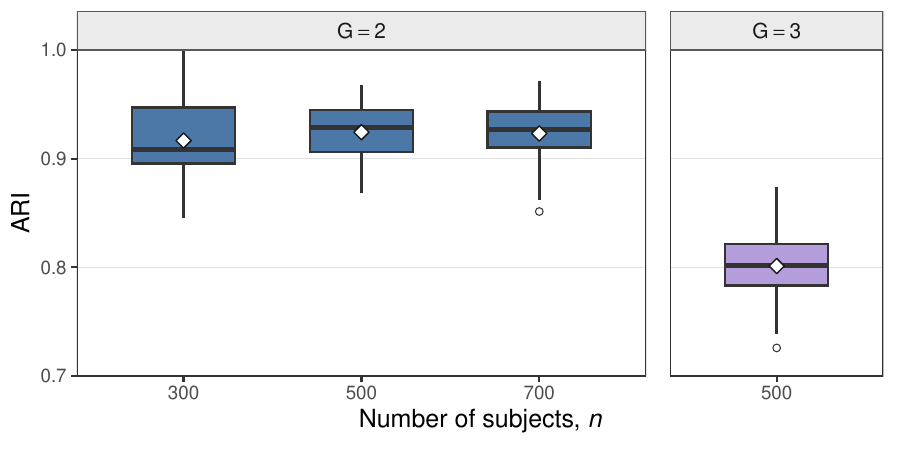}
\end{minipage}

\caption{
\textbf{Parameter estimation and latent-class recovery in the simulation studies.}
\textbf{(A)} Mean relative $L_2$ errors for the fixed-effect coefficients and relative Frobenius errors for the covariance matrices in the $G=2$ setting. Points denote means across 100 replications, and error
bars represent $\pm 1$ standard deviation.
\textbf{(B)} Distributions of the ARI across 100
replications for the $G=2$ and $G=3$ settings. White diamonds on the boxplots denote means.
}
\label{fig:sim_param_est}
\end{figure}

Parameter recovery was evaluated using the relative $L_2$ error for the vectorized fixed-effect coefficients $\boldsymbol{\beta}_g$ and the relative Frobenius error for the covariance matrices $\boldsymbol{\Omega}_g$ and $\boldsymbol{\Sigma}_g$. Specifically, for replication $b$, these errors were defined as
$\operatorname{RE}_{\boldsymbol{\beta}_g}^{(b)}
= \|\operatorname{vec}(\widehat{\boldsymbol{\beta}}_g^{(b)}
- \boldsymbol{\beta}_g)\|_2/
\|\operatorname{vec}(\boldsymbol{\beta}_g)\|_2$
and
$\operatorname{RE}_{\mathbf{A}_g}^{(b)}
= \|\widehat{\mathbf{A}}_g^{(b)}-\mathbf{A}_g\|_{\mathrm{F}}/
\|\mathbf{A}_g\|_{\mathrm{F}}$, respectively, where
$\mathbf{A}_g\in\{\boldsymbol{\Omega}_g,\boldsymbol{\Sigma}_g\}$.
For $\boldsymbol{\Omega}_g$, the norms were computed using only its structurally free within-outcome blocks. Figure~\ref{fig:sim_param_est} A presents the mean relative errors across 100 replications, with error bars representing one standard deviation. Clearly, both the means and standard deviations decreased consistently as $n$ increased. Although the random-effect covariance matrices were the most difficult parameter blocks to estimate, their errors also declined substantially with increasing sample size.

Figure~\ref{fig:sim_param_est} B presents boxplots of the ARI values, quantifying latent-class recovery, across 100 replications for the three $G=2$ sample sizes and the $G=3, n=500$ setting. Recovery was consistently strong in the $G=2$ setting, with median ARIs of $0.909$, $0.929$, and $0.927$ for $n=300$, $500$, and $700$, respectively. In the more challenging $G=3$ setting, the median ARI was $0.802$, with the central 50\% of ARIs ranging from $0.783$ to $0.821$ ($\mathrm{IQR} = 0.038$), indicating substantial and stable agreement with the true clustering despite the weak trajectory separation. We also assessed the selection of the number of latent classes using the variational BIC. Candidate models with $G\in \{1,2,3\}$ were fitted to data generated with $G=2$, whereas models with $G\in\{1,2,3,4\}$ were fitted to data generated with $G=3$. The variational BIC selected the true number of latent classes in all replications across both settings.

The $G=1$ experiments likewise yielded accurate parameter estimates, with estimation accuracy improving as $n$ increased. Detailed results for $G=1$ setting, numerical summaries for all parameter blocks, mean-absolute-error heatmaps for covariance matrices, and additional latent-class recovery results are provided in the Supplementary Material. 

\subsection{Comparison with existing methods}
\label{subsec:simulations2}
We compared the proposed VEM with existing approaches according to their overlapping modeling capabilities. Table~\ref{tab:method_capabilities} summarizes the modelling capabilities of the proposed method and some published methods for longitudinal data. For the homogeneous setting ($G=1$), parameter estimates were compared with \texttt{lme4} \citep{bates2015lme4}; because this setting does not involve latent-class recovery, its detailed results are provided in the Supplementary Material. For the heterogeneous settings ($G=2$ and $G=3$), we compared latent-class recovery with \texttt{lcmm} \citep{proust2017estimation} and the Bayesian clusterwise multivariate generalized linear mixed-model approach of \citet{vavra2024clusterwise}, hereafter referred to as \texttt{clustGLMM}. Because the methods differ in their model specifications, comparisons focused on latent-class recovery and accuracy in selecting the true number of latent classes $G$. 

Compared with \texttt{lcmm}, the proposed method consistently achieved higher
median ARIs across all four heterogeneous settings. For $G=2$,
the median ARIs from \texttt{lcmm} were 0.883, 0.891, and
0.883 for $n=300$, $500$, and $700$, respectively, compared
with 0.909, 0.929, and 0.927 for the proposed method.
In the more challenging $G=3$, $n=500$ setting, performance
was similar, with median ARIs of 0.798 for \texttt{lcmm}
and 0.802 for the proposed method. The proposed method also showed slightly higher quartiles and a narrower IQR than \texttt{lcmm} (0.783--0.821 versus 0.774--0.818), suggesting modestly improved latent-class recovery and stability. Both the proposed method using variational BIC and \texttt{lcmm} using BIC selected the true number of classes in all replications of each setting. A key distinction is that the \texttt{lcmm} benchmark used Gaussian models for log-transformed responses, whereas our proposed method operates directly on the raw counts. This formulation respects the discrete, nonnegative nature of the responses and explicitly accommodates over-dispersion without requiring a response transformation.

The comparison with \texttt{clustGLMM} yielded a markedly different pattern. Its sparse finite-mixture procedure specifies an upper bound $G_{\max}$ and estimates the number of classes as the posterior mode of the number of occupied components over the retained MCMC draws, rather than by fitting separate candidate models and comparing an information criterion. With $G_{\max}=3$ for the $G=2$ settings and $G_{\max}=4$ for the $G=3$ setting, \texttt{clustGLMM} selected the upper bound in all 100 replications. Consequently, the median ARIs of the selected \texttt{clustGLMM} partitions were 0.444, 0.448, and 0.452 for the true $G=2$ simulation at $n=300$, $500$, and $700$, respectively, and 0.264 for the true $G=3$, $n=500$ simulation. This consistent over-partitioning may reflect both the inability to model over-dispersion, the occupied-component selection procedure, and differences between the fitted and generating models: \texttt{clustGLMM} assumes constant mixing proportions and does not include the visit-level multivariate Gaussian error used in the proposed model, so additional components may accommodate covariate-dependent class-membership patterns or residual heterogeneity. Thus, for these simulation settings, the proposed method recovered both the number of classes and the generating partition more accurately than \texttt{clustGLMM}.

\begin{table*}[!t]
\centering
\caption{
Comparison of the modelling capabilities of existing methods and the
proposed latent-class Poisson-lognormal mixed model.
}
\label{tab:method_capabilities}

\footnotesize
\setlength{\tabcolsep}{2.8pt}
\renewcommand{\arraystretch}{1.06}

\begin{tabularx}{
  \textwidth
}{
  @{}
  >{\raggedright\arraybackslash}p{0.255\textwidth}
  *{5}{>{\centering\arraybackslash}X}
  @{}
}
\toprule
\textbf{Modelling capability}
& \texttt{lcmm}
& \texttt{lme4}
& \texttt{longclust}
& \texttt{clustGLMM}
& Proposed method \\
\midrule

Native Poisson likelihood
& No
& Yes
& No
& Yes\textsuperscript{1}
& Yes \\

Extra-Poisson variation
& N/A
& Yes\textsuperscript{2}
& N/A
& Yes\textsuperscript{1}
& Yes \\

Multiple longitudinal responses
& Yes\textsuperscript{3}
& Via stacking\textsuperscript{2}
& No\textsuperscript{4}
& Yes
& Yes \\

General fixed-effect covariates
& Yes
& Yes
& No\textsuperscript{4}
& Yes
& Yes \\

Cross-response dependence
& Yes\textsuperscript{3}
& Via stacking\textsuperscript{2}
& N/A
& Yes\textsuperscript{1}
& Yes \\

Latent-class analysis
& Yes
& No
& Yes
& Yes
& Yes \\

Covariate-dependent class membership
& Yes
& N/A
& No
& No
& Yes \\

Irregular subject-specific visits
& Yes
& Yes
& No\textsuperscript{4}
& Yes
& Yes \\

Missing responses
& Observed-data likelihood (MAR)
& Observed-data likelihood (MAR)\textsuperscript{5}
& No\textsuperscript{4}
& Data augmentation\textsuperscript{6}
& Observed-data likelihood (MAR) \\

\bottomrule
\end{tabularx}

\vspace{0.25em}

\begin{minipage}{\textwidth}
\scriptsize
\raggedright
\textit{Notes:}
\textsuperscript{1}
The implementation accompanying \texttt{clustGLMM} includes Poisson
log-linear mixed models. Gaussian random effects induce marginal
over-dispersion and may be correlated across outcomes.
\textsuperscript{2}
\texttt{lme4} has no native multivariate-response interface. In our
one-component benchmark, outcomes are stacked in long format, with
outcome-specific fixed and random effects. Random effects induce
extra-Poisson variation and may represent cross-response dependence.
\textsuperscript{3}
\texttt{lcmm} accommodates multiple outcomes and induces their dependence
through a shared latent process.
\textsuperscript{4}
\texttt{longclust} requires a complete rectangular matrix and supports
selected temporal mean structures, but not general covariate-dependent
regression, distinct multivariate-response processes, or
covariate-dependent class membership.
\textsuperscript{5}
In the stacked implementation, a row with a missing response is omitted;
other observed responses and visits from that subject remain in the
likelihood. This is an observed-data likelihood analysis under MAR.
\textsuperscript{6}
\texttt{clustGLMM} accommodates missing outcomes through Bayesian data
augmentation when the corresponding covariates are observed.
\end{minipage}

\end{table*}

\section{Real data application: longitudinal COVID-19 neutrophil RNA-seq data}
\label{sec:real_data}
We applied the proposed model to longitudinal RNA-seq data from a study of circulating neutrophils in patients presenting with suspected COVID-19 \citep{lasalle2022longitudinal}. The original study included 384 adults who presented to the Massachusetts General Hospital emergency department with acute respiratory
symptoms between March and May 2020, of whom 306 tested positive for SARS-CoV-2 and 78 were symptomatic test-negative controls. Peripheral blood samples were collected at hospital admission (day 0) and, for patients who
remained hospitalized, on days 3 and 7. Neutrophils were enriched from whole blood by negative selection and profiled using bulk RNA sequencing. The processed RNA-seq count data are publicly available through the Gene Expression Omnibus under accession GSE212041.

Starting from 781 sequenced samples, we applied the six sample-level RNA-seq quality-control criteria used by \cite{lasalle2022longitudinal} and restricted the analysis to scheduled longitudinal visits at days 0, 3, and 7, resulting in a final dataset of 662 RNA-seq samples from 364 subjects. Note that study participants retained in the analysis dataset may or may not have had measurements at all three visits. 

Clinical severity was defined according to the most severe disease status observed within 28 days of enrollment \citep{lasalle2022longitudinal}. Patients who died
within 28 days or required mechanical ventilation were classified as severe, whereas those requiring supplemental oxygen, those hospitalized without supplemental oxygen, and those discharged without hospitalization were classified as non-severe. The original study identified substantial
longitudinal heterogeneity in neutrophil transcriptional states and demonstrated that several of these states were associated with clinical severity. This longitudinal heterogeneity makes the dataset suitable for
investigating whether our proposed model can identify clinically meaningful subgroups. 

Motivated by the transcriptional patterns reported by \cite{lasalle2022longitudinal}, we considered \textit{ARG1}, \textit{SERPINB2}, and \textit{ZBTB16} as the multivariate count responses. The original study associated \textit{ARG1} with neutrophil degranulation and identified distinct longitudinal severity-related patterns for \textit{SERPINB2} and \textit{ZBTB16}. Following quality control, trimmed mean of M-values (TMM)-based normalization factors were estimated in edgeR \citep{robinson2010scaling, ChenEtAl2025}. The original RNA-seq counts were modeled using the effective library size, the library size times its TMM normalization factor, as an offset.
\begin{equation}
o_{il}=\log(\text{library size}_{il}\times\text{normalization factor}_{il}).
\label{eq:real_offset}
\end{equation}

We fitted nine candidate models: the one-component model was fitted without a gating network; for each $G\in \{2,3,4,5\}$, we considered both an intercept-only gating network and a gating network containing the centered ordinal Age variable. In every model, the fixed effects included an intercept, time (scaled as day divided by seven), and centered Age. The subject-specific random effects comprised a random intercept and random time slope.
For each model with $G>1$, five independently seeded $k$-means-based initializations were considered, with $k$-means run using 50 random starts. The fit attaining the largest final ELBO among the five runs was retained. The candidate models were then compared using the variational BIC. After examination of the convergence diagnostics, the two-component model with Age in the gating network attained the smallest variational BIC and was selected for subsequent interpretation. 

Maximum \textit{a posteriori} classification assigned 105 subjects to Class~1 and 259 subjects to Class~2. In the estimated gating network, the coefficient of centered Age for the log-odds of Class~1 relative to Class~2 was $-0.6720$, corresponding to an odds ratio of $0.5107$.
Thus, each one-category increase in Age was associated with an approximately $49\%$ reduction in the odds of Class~1 versus Class~2 membership. Consistently, the mean ordinal age-category score was $2.4381$ in Class~1 and $3.4981$ in Class~2. The corresponding median categories were category 2 (35--49 years) and category 4 (65--79 years), respectively, indicating that Class~2 contained an older subject population.

\begin{figure}[t]
\centering

\begin{minipage}{0.7\textwidth}
\textbf{A}\par\vspace{-0.1em}
\centering
\includegraphics[
    width=\textwidth 
]{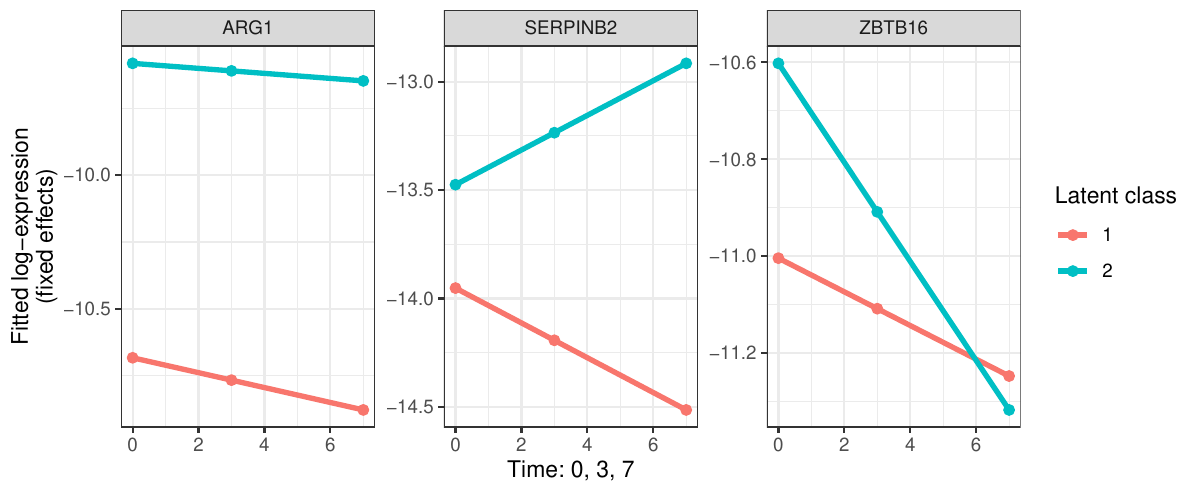}
\end{minipage}

\vspace{0.05em}

\begin{minipage}{0.7\textwidth}
\textbf{B}\par\vspace{-0.1em}
\centering
\includegraphics[
    width=\textwidth 
]{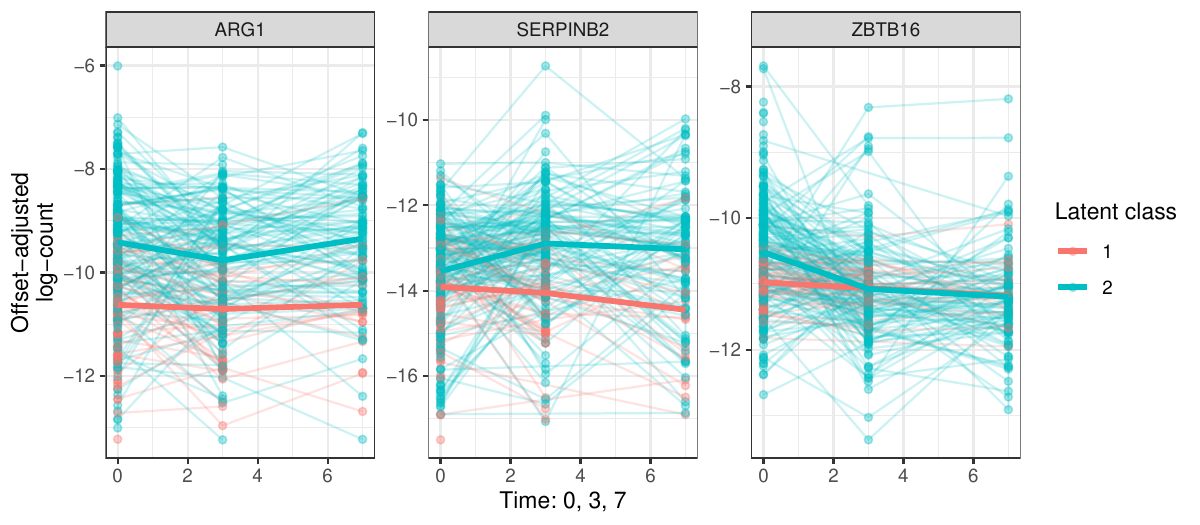}
\end{minipage}

\caption{
\textbf{Estimated and observed longitudinal expression trajectories for the three selected genes.}
\textbf{(A)} Class-specific fitted latent log-expression trajectories for \textit{ARG1}, \textit{SERPINB2}, and \textit{ZBTB16}, evaluated at the mean age of the analyzed subjects. \textbf{(B)} Observed offset-adjusted log-count trajectories, defined for visualization as $\log(Y_{ijl}+0.5)-o_{il}$, where the pseudocount 0.5 avoids taking the logarithm of zero and $o_{il}$ is defined in Equation~\eqref{eq:real_offset}.
Thin lines represent individual subject trajectories colored according to their maximum-\textit{a-posteriori} latent-class assignments, whereas thick lines show the corresponding class-specific mean trajectories.
}
\label{fig:covid-expression-trajectories}
\end{figure}

Figure~\ref{fig:covid-expression-trajectories} compares the estimated class-specific fixed-effect trajectories with the observed offset-adjusted log counts. The fitted trajectories in Figure~\ref{fig:covid-expression-trajectories} A broadly reproduce the empirical class-specific patterns in Figure~\ref{fig:covid-expression-trajectories} B. For \textit{ARG1}, Class~2 exhibited consistently higher expression than Class~1, while temporal changes within each class were comparatively modest. For \textit{SERPINB2}, the fitted expression decreased over time in Class~1 but increased in Class~2. The observed Class~2 mean displayed a pronounced increase at day 3 and remained higher at day 7 than at admission. For \textit{ZBTB16}, expression declined in both classes, with a substantially steeper decline in Class~2.
Thus, the fitted trajectories were qualitatively consistent with the observed class-specific averages. Some observed means exhibited modest nonlinear movement at day 3, particularly for \textit{ARG1} and \textit{SERPINB2}, which was smoothed by the linear time specification. Exact agreement is not expected because the fitted curves represent fixed-effect trajectories at centered Age, whereas the observed summaries are unadjusted for Age and contain subject-specific and residual variation.

\begin{figure}[t]
\centering

\includegraphics[
    width=0.8\columnwidth,
    keepaspectratio
]{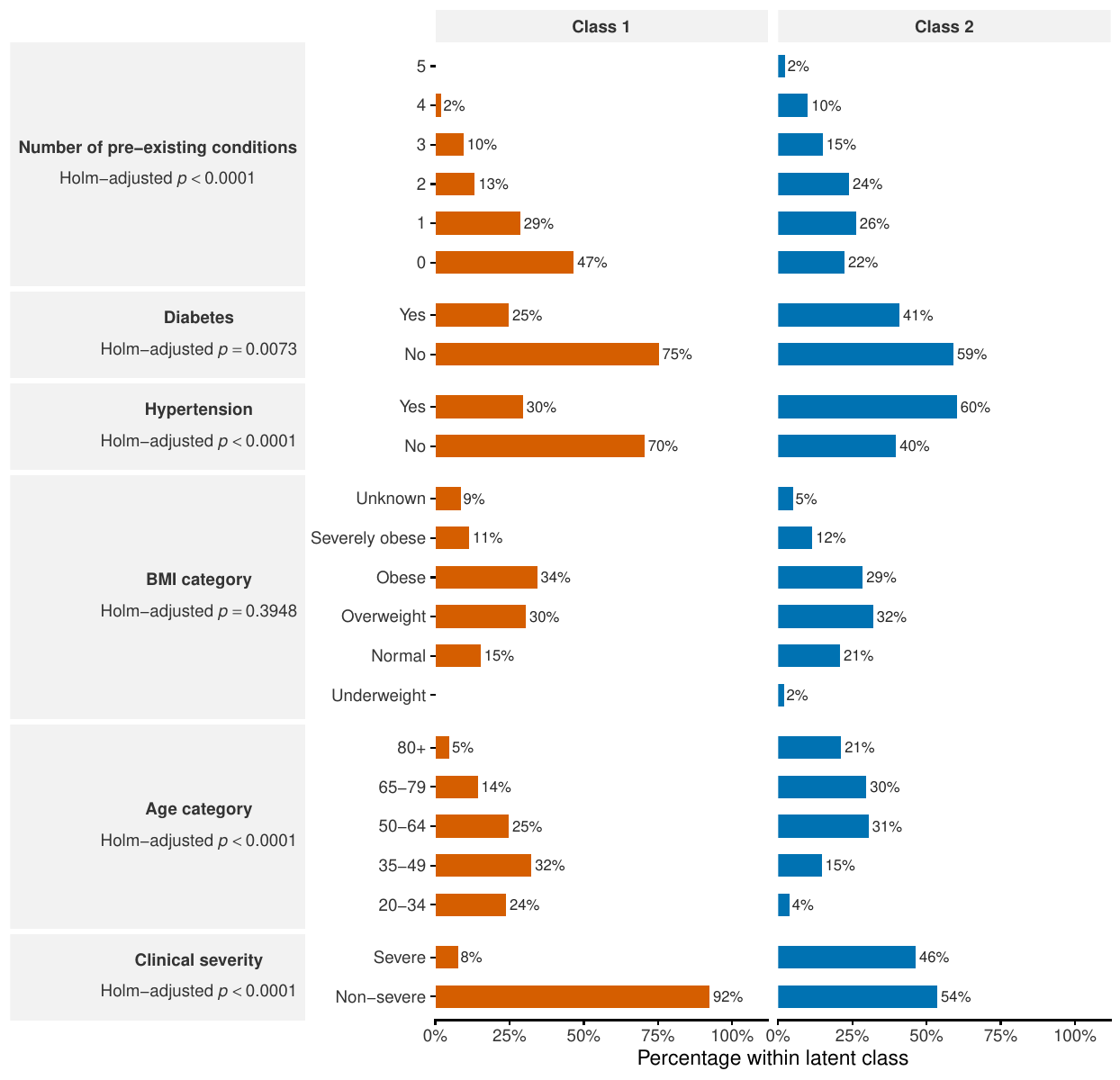}

\caption{
Characteristics of the two inferred latent classes.
Horizontal bars show the percentage of subjects in each category, calculated separately within each class among subjects with observed data.
The Holm-adjusted $p$-value from a Pearson chi-squared test of homogeneity comparing the  categorical distributions between the two classes is also provided.
}
\label{fig:real_barplot_pvalue}

\end{figure}

Furthermore, the two inferred classes exhibited markedly different clinical profiles (Table~\ref{tab:real_binary_summary} and Figure~\ref{fig:real_barplot_pvalue}). Consistent with the estimated Age effect in the gating network, around 51\% of Class~2 subjects were aged 65 years or older, compared with 19\% of Class~1 subjects. Importantly, clinical severity and the comorbidity variables were not included in model fitting. Nevertheless, compared with Class~1, Class~2 contained substantially larger proportions of subjects with severe clinical status, 
hypertension, 
diabetes, 
and at least two pre-existing conditions among heart disease, lung disease, kidney disease, diabetes, and hypertension. 
Conversely, subjects with no recorded pre-existing conditions were more prevalent in Class~1.

\begin{table}[t]
\centering
\caption{Selected clinical characteristics of the two latent classes. Percentages were calculated within class among subjects with observed data.}
\label{tab:real_binary_summary}

\normalsize
\setlength{\tabcolsep}{3pt}

\begin{tabular}{@{}lcc@{}}
\toprule
Characteristic & \shortstack{Class 1\\($n=105$)} & \shortstack{Class 2\\($n=259$)} \\
\midrule
Age $\geq 65$ years & 19\% & 51\% \\
Severe clinical status & 8\% & 46\% \\
Hypertension & 30\% & 60\% \\
Diabetes & 25\% & 41\% \\
$\geq 2$ pre-existing conditions & 25\% & 51\% \\
No pre-existing conditions & 47\% & 22\% \\
\bottomrule
\end{tabular}
\end{table}

Pearson chi-squared tests of homogeneity were used to compare the six clinical characteristics between the two latent classes, with $p$-values adjusted using Holm's procedure \citep{holm1979simple}. BMI values recorded as ``Unknown'' were treated as missing. Significant between-class differences were detected for Age category, clinical severity, hypertension, diabetes, and the number of pre-existing conditions (Figure~\ref{fig:real_barplot_pvalue}) but not for BMI category. 
Thus, although clinical severity and the comorbidity variables were not used to define the latent classes, the classes inferred from the multivariate longitudinal gene-expression trajectories exhibited distinct and clinically interpretable profiles.

The joint trajectory and clinical results are also consistent with \citet{lasalle2022longitudinal}. In that study, \textit{ARG1} was enriched in severe patients across sampling days, \textit{SERPINB2} increased over time in severe patients but decreased slightly in non-severe patients, and \textit{ZBTB16} was more highly expressed at admission in severe patients and subsequently declined. Correspondingly, as shown in Figure~\ref{fig:covid-expression-trajectories} A, our clinically more severe Class~2 exhibited higher \textit{ARG1}, an increasing \textit{SERPINB2} trajectory, and higher initial but more rapidly decreasing \textit{ZBTB16}. 

Overall, this illustrative application highlights the ability of the proposed model to characterize heterogeneity in multivariate longitudinal count trajectories and to identify clinically interpretable latent subgroups. 

\section{Discussion}
\label{sec:discussion}

We introduced a latent-class Poisson-lognormal mixed model for clustering subjects according to heterogeneous multivariate longitudinal count trajectories. The framework jointly 
accounts for over-dispersed count data, serial dependence, cross-outcome dependence, subject-specific observation times, and covariate-dependent class membership. The proposed VEM algorithm makes estimation feasible without repeatedly performing high-dimensional numerical integration or MCMC sampling. Simulations demonstrated that the proposed procedure yielded accurate parameter estimates, 
and achieved strong latent-class recovery under limited separation between class-specific trajectories, and reliably selected the true number of latent classes using the variational BIC. While the simulations supported variational BIC as a practical criterion for selecting the number of latent classes,  it remains an approximation to the classical BIC and may be affected by the tightness of the ELBO and local optima. We therefore used multiple initializations and retained the fit with the largest final ELBO before interpreting the selected model. When more accurate latent-posterior inference is of primary interest, a hybrid strategy combining an initial VGA-based VEM fit with subsequent class-conditional MCMC refinement provides a possible alternative \citep{silva2023finite, fang2023clustering}.

In the neutrophil RNA-seq application, the selected two-class model identified distinct trajectories for \textit{ARG1}, \textit{SERPINB2}, and \textit{ZBTB16}. The inferred classes also differed substantially in age, clinical severity, and comorbidity burden, although clinical severity and the comorbidity variables were not included in model fitting. While these post hoc associations should be regarded as descriptive rather than causal, their agreement with patterns reported by \cite{lasalle2022longitudinal} provides additional biological interpretability. 

Two directions for future work are particularly promising. First, extending the proposed framework to high-dimensional multivariate responses would broaden its applicability to larger genomic panels. Factor-analyzer techniques provide a natural approach by representing high-dimensional dependence through a lower-dimensional latent structure, thereby reducing the number of covariance parameters \citep{mclachlan2003modelling, payne2025finite}. Second, the current model assumes that each subject retains the same latent-class membership over time. A dynamic extension could allow subjects to transition between latent classes over time and permit class prevalence to evolve, allowing latent states to emerge, diminish, or re-emerge over time. Such extensions would provide greater flexibility while retaining the longitudinal clustering perspective developed here.

Overall, the proposed framework provides a flexible and computationally tractable approach for clustering complex multivariate longitudinal count data. The simulation and real-data results demonstrate its potential to recover meaningful trajectory subgroups while accounting for the dependence and over-dispersion inherent in longitudinal biological count data.


\clearpage
\etocdepthtag.toc{supplement}

\setcounter{section}{0}
\setcounter{equation}{0}
\setcounter{figure}{0}
\setcounter{table}{0}

\renewcommand{\thesection}{S\arabic{section}}
\renewcommand{\thesubsection}{\thesection.\arabic{subsection}}
\renewcommand{\thesubsubsection}{\thesubsection.\arabic{subsubsection}}

\renewcommand{\theequation}{S\arabic{equation}}
\renewcommand{\thefigure}{S\arabic{figure}}
\renewcommand{\thetable}{S\arabic{table}}

\renewcommand{\theHsection}{supp.\arabic{section}}
\renewcommand{\theHequation}{supp.\arabic{equation}}
\renewcommand{\theHfigure}{supp.\arabic{figure}}
\renewcommand{\theHtable}{supp.\arabic{table}}

\pdfbookmark[0]{Supplementary Material}{supplement-title}

\begin{center}
  {\LARGE\bfseries Supplementary Material\par}

  \vspace{0.6em}

  {\large\itshape
  Latent-Class Poisson-Lognormal Mixed Models for Clustering \\
  Multivariate Longitudinal Count Data\par}
\end{center}

\vspace{1em}

\begingroup
  \etocsettagdepth{main}{none}
  \etocsettagdepth{supplement}{subsection}
  \etocsetnexttocdepth{subsection}
  \renewcommand{\contentsname}{Supplementary contents}
  \tableofcontents
\endgroup

\section{Variational EM algorithm and implementation details}
\label{sec:supp-vem}

This section gives the estimation details of the proposed VEM procedure. As defined in the main manuscript, unequal numbers and irregular timing of visits are represented by the subject-specific \(T_i\) and \(t_{il}\).

\subsection{Stacked notation and Gaussian representation}
\label{subsec:supp-stacked}

Suppose that each fixed-effect design vector
\(\boldsymbol{X}_{ij}(t_{il})\) has dimension \(q\), and each random-effect
design vector \(\boldsymbol{R}_{ij}(t_{il})\) has dimension \(r\). Define
\begin{align*}
 \boldsymbol{Y}_i
 &=\left(\boldsymbol{Y}_{i1}^{\trans},\ldots,
          \boldsymbol{Y}_{iT_i}^{\trans}\right)^{\trans}
   \in\R^{pT_i},\\
 \boldsymbol{\eta}_{ig}
 &=\left(\boldsymbol{\eta}_{i1g}^{\trans},\ldots,
          \boldsymbol{\eta}_{iT_i g}^{\trans}\right)^{\trans}
   \in\R^{pT_i},\\
 \boldsymbol{b}_{ig}
 &=\left(\boldsymbol{b}_{i1g}^{\trans},\ldots,
          \boldsymbol{b}_{ipg}^{\trans}\right)^{\trans}
   \in\R^{pr},\\
 \boldsymbol{\beta}_g
 &=\left(\boldsymbol{\beta}_{g1}^{\trans},\ldots,
          \boldsymbol{\beta}_{gp}^{\trans}\right)^{\trans}
   \in\R^{pq}.
\end{align*}
The entries of \(\boldsymbol{Y}_i\) and \(\boldsymbol{\eta}_{ig}\) are ordered
by visit and then by outcome, so the position of outcome \(j\) at visit \(l\) is
\(k(j,l):=(l-1)p+j\).

Let \(\boldsymbol{X}_i\in\R^{pT_i\times pq}\) be the stacked fixed-effect
design matrix: row \(k(j,l)\) contains
\(\boldsymbol{X}_{ij}(t_{il})^{\trans}\) in the columns associated with
\(\boldsymbol{\beta}_{gj}\) and zeros elsewhere. Likewise, let
\(\boldsymbol{M}_{il}\in\R^{p\times pr}\) have
\(\boldsymbol{R}_{ij}(t_{il})^{\trans}\) in the outcome-\(j\) random-effect
block of row \(j\), and define
\[
 \boldsymbol{M}_i
 =\left(\boldsymbol{M}_{i1}^{\trans},\ldots,
        \boldsymbol{M}_{iT_i}^{\trans}\right)^{\trans}.
\]
The known stacked offset is
\begin{equation*}
 \boldsymbol{o}_i
 =\left(o_{i1}\one_p^{\trans},\ldots,
        o_{iT_i}\one_p^{\trans}\right)^{\trans}
 \in\R^{pT_i}.
\end{equation*}
Conditional on \(Z_i=g\), the Gaussian part of the model can then be written
as
\begin{align}
 \boldsymbol{b}_{ig}
 &\sim N(\boldsymbol{0},\boldsymbol{\Omega}_g),\nonumber\\
 \boldsymbol{\eta}_{ig}\mid\boldsymbol{b}_{ig},Z_i=g
 &\sim N\!\left(
      \boldsymbol{X}_i\boldsymbol{\beta}_g+
      \boldsymbol{M}_i\boldsymbol{b}_{ig}+\boldsymbol{o}_i,
      \boldsymbol{K}_{\Sigma,ig}\right),
 \qquad
 \boldsymbol{K}_{\Sigma,ig}
 =\boldsymbol{I}_{T_i}\otimes\boldsymbol{\Sigma}_g,
 \label{eq:supp-stacked-model}
\end{align}
with
\(\boldsymbol{\Omega}_g=\blockdiag\{\boldsymbol{\Omega}_g^{(1)},
\ldots,\boldsymbol{\Omega}_g^{(p)}\}\). 

\subsection{Flexible variational family and component lower bound}
\label{subsec:supp-family-elbo}

Conditional on \(Z_i=g\), the variational family
used in the main manuscript is
\begin{align}
 q_{ig}(\boldsymbol{b}_{ig},\boldsymbol{\eta}_{ig})
 &=q_{ig}(\boldsymbol{b}_{ig})
   q_{ig}(\boldsymbol{\eta}_{ig}\mid\boldsymbol{b}_{ig}),\nonumber\\
 q_{ig}(\boldsymbol{b}_{ig})
 &=N\!\left(\boldsymbol{\mu}_{ig}^{(b)},
             \boldsymbol{S}_{ig}^{(b)}\right),\nonumber\\
 q_{ig}(\boldsymbol{\eta}_{ig}\mid\boldsymbol{b}_{ig})
 &=N\!\left(\boldsymbol{A}_{ig}+
             \boldsymbol{L}_{ig}\boldsymbol{b}_{ig},
             \boldsymbol{V}_{ig}\right).
 \label{eq:supp-var-family}
\end{align}
Here \(\boldsymbol{S}_{ig}^{(b)}\) and \(\boldsymbol{V}_{ig}\) may be full,
and \(\boldsymbol{L}_{ig}\) is unrestricted. These are properties of the
variational posterior; they do not alter the block-diagonal model constraint
on \(\boldsymbol{\Omega}_g\) or the model covariance
\(\boldsymbol{I}_{T_i}\otimes\boldsymbol{\Sigma}_g\). By using this flexible variational family, the algorithm can obtain a more accurate approximation to the conditional posterior of $(\boldsymbol{b}_{ig}, \boldsymbol{\eta}_{ig})$. Furthermore, let \(\tau_{ig}=q_i(Z_i=g)\) denote the variational posterior class-membership probability.

The marginal variational moments of \(\boldsymbol{\eta}_{ig}\) are
\begin{align*}
 \widetilde{\boldsymbol{\mu}}_{ig}
 &=\boldsymbol{A}_{ig}
   +\boldsymbol{L}_{ig}\boldsymbol{\mu}_{ig}^{(b)},\nonumber\\
 \widetilde{\boldsymbol{V}}_{ig}
 &=\boldsymbol{V}_{ig}
   +\boldsymbol{L}_{ig}\boldsymbol{S}_{ig}^{(b)}
    \boldsymbol{L}_{ig}^{\trans}.
\end{align*}
Write \(\widetilde{\mu}_{ijlg}\) for entry \(k(j,l)\) of
\(\widetilde{\boldsymbol{\mu}}_{ig}\), and
\(\widetilde{v}_{ijlg}\) for the corresponding diagonal entry of
\(\widetilde{\boldsymbol{V}}_{ig}\). Then
\begin{equation*}
 \lambda_{ijlg}
 :=\mathbb{E}_{q_{ig}}\{\exp(\eta_{ijlg})\}
 =\exp\!\left(\widetilde{\mu}_{ijlg}
        +\frac{1}{2}\widetilde{v}_{ijlg}\right).
\end{equation*}

For later use, define
\begin{align}
 \boldsymbol{d}_{ig}
 &=\boldsymbol{A}_{ig}-\boldsymbol{o}_i
   -\boldsymbol{X}_i\boldsymbol{\beta}_g
   +(\boldsymbol{L}_{ig}-\boldsymbol{M}_i)
     \boldsymbol{\mu}_{ig}^{(b)},\nonumber\\
 \boldsymbol{\Psi}_{ig}
 &=\boldsymbol{V}_{ig}
   +(\boldsymbol{L}_{ig}-\boldsymbol{M}_i)
     \boldsymbol{S}_{ig}^{(b)}
     (\boldsymbol{L}_{ig}-\boldsymbol{M}_i)^{\trans}
   +\boldsymbol{d}_{ig}\boldsymbol{d}_{ig}^{\trans}.
 \label{eq:supp-Psi}
\end{align}
Thus, \(\boldsymbol{\Psi}_{ig}\) is the variational expectation of the outer
product of
\(\boldsymbol{\eta}_{ig}-\boldsymbol{o}_i-
\boldsymbol{X}_i\boldsymbol{\beta}_g-
\boldsymbol{M}_i\boldsymbol{b}_{ig}\).

Let \(d_b=pr\) and \(d_{\eta,i}=pT_i\). After cancellation of the Gaussian
normalizing constants, the complete component lower bound is
\begin{align}
 B_{ig}
 ={}&-\frac{1}{2}\log|\boldsymbol{\Omega}_g|
 -\frac{1}{2}\tr\!\left[
   \boldsymbol{\Omega}_g^{-1}
   \left\{\boldsymbol{S}_{ig}^{(b)}+
   \boldsymbol{\mu}_{ig}^{(b)}
   \boldsymbol{\mu}_{ig}^{(b)\trans}\right\}\right]
 +\frac{1}{2}\log|\boldsymbol{S}_{ig}^{(b)}|
 +\frac{d_b}{2}\nonumber\\
 &-\frac{T_i}{2}\log|\boldsymbol{\Sigma}_g|
 -\frac{1}{2}\tr\!\left(
   \boldsymbol{K}_{\Sigma,ig}^{-1}\boldsymbol{\Psi}_{ig}\right)
 +\frac{1}{2}\log|\boldsymbol{V}_{ig}|
 +\frac{d_{\eta,i}}{2}\nonumber\\
 &+\sum_{l=1}^{T_i}\sum_{j=1}^{p}
 \left\{Y_{ijl}\widetilde{\mu}_{ijlg}
 -\lambda_{ijlg}-\log\Gamma(Y_{ijl}+1)\right\}.
 \label{eq:supp-component-elbo}
\end{align}
The mixture ELBO is
\begin{equation}
 \mathcal{L}(q,\boldsymbol{\theta})
 =\sum_{i=1}^{n}\sum_{g=1}^{G}
 \tau_{ig}\{\log\pi_{ig}-\log\tau_{ig}+B_{ig}\}.
 \label{eq:supp-mixture-elbo}
\end{equation}

\subsection{V-step through a joint Gaussian parameterization}
\label{subsec:supp-vstep}

For fixed model parameters, the V-step is most conveniently performed with
the joint latent vector
\[
 \boldsymbol{U}_{ig}
 =\left(\boldsymbol{b}_{ig}^{\trans},
        \boldsymbol{\eta}_{ig}^{\trans}\right)^{\trans},
 \qquad
 q_{ig}(\boldsymbol{U}_{ig})
 =N(\boldsymbol{m}_{ig},\boldsymbol{C}_{ig}).
\]
Partition its mean and covariance as
\begin{equation*}
 \boldsymbol{m}_{ig}
 =\begin{pmatrix}\boldsymbol{m}_{ig,b}\\
                  \boldsymbol{m}_{ig,\eta}\end{pmatrix},
 \qquad
 \boldsymbol{C}_{ig}
 =\begin{pmatrix}
   \boldsymbol{C}_{ig,bb}&\boldsymbol{C}_{ig,b\eta}\\
   \boldsymbol{C}_{ig,\eta b}&\boldsymbol{C}_{ig,\eta\eta}
  \end{pmatrix}.
\end{equation*}
The Gaussian model in equation~\eqref{eq:supp-stacked-model} implies the prior
mean
\begin{equation}
 \boldsymbol{m}_{ig,0}
 =\begin{pmatrix}
   \boldsymbol{0}\\
   \boldsymbol{X}_i\boldsymbol{\beta}_g+\boldsymbol{o}_i
  \end{pmatrix}
 \label{eq:supp-prior-mean}
\end{equation}
and the prior covariance matrix
\begin{equation*}
\boldsymbol{P}_{ig}
= \text{Var}(\boldsymbol{U}_{ig} \mid Z_i = g)
 =\begin{pmatrix}
 \boldsymbol{\Omega}_g
 &\boldsymbol{\Omega}_g \boldsymbol{M}_i^\top\\
 \boldsymbol{M}_i \boldsymbol{\Omega}_g
 &\boldsymbol{M}_i\boldsymbol{\Omega}_g \boldsymbol{M}_i^\top + \boldsymbol{K}_{\Sigma, ig}
 \end{pmatrix}, 
\end{equation*}
so the prior precision matrix is
\begin{equation}
 \boldsymbol{Q}_{ig}
 = \boldsymbol{P}_{ig}^{-1}
 =\begin{pmatrix}
 \boldsymbol{\Omega}_g^{-1}
 +\boldsymbol{M}_i^{\trans}\boldsymbol{K}_{\Sigma,ig}^{-1}
  \boldsymbol{M}_i
 &-\boldsymbol{M}_i^{\trans}\boldsymbol{K}_{\Sigma,ig}^{-1}\\
 -\boldsymbol{K}_{\Sigma,ig}^{-1}\boldsymbol{M}_i
 &\boldsymbol{K}_{\Sigma,ig}^{-1}
 \end{pmatrix}.
 \label{eq:supp-prior-precision}
\end{equation}

Let \(\boldsymbol{\lambda}_{ig}\in\R^{pT_i}\) have entries
\begin{equation}
 (\boldsymbol{\lambda}_{ig})_{k(j,l)}
 =\exp\!\left\{(\boldsymbol{m}_{ig,\eta})_{k(j,l)}
 +\frac{1}{2}(\boldsymbol{C}_{ig,\eta\eta})_{k(j,l),k(j,l)}\right\},
 \label{eq:supp-joint-lambda}
\end{equation}
and define
\[
 \boldsymbol{D}_{ig}
 =\blockdiag\{\boldsymbol{0}_{pr\times pr},
               \diag(\boldsymbol{\lambda}_{ig})\}.
\]
Holding \(\boldsymbol{C}_{ig}\) fixed, the gradient and Hessian of \(B_{ig}\)
with respect to \(\boldsymbol{m}_{ig}\) are
\begin{align}
 \boldsymbol{g}_{ig}
 &=-\boldsymbol{Q}_{ig}
   (\boldsymbol{m}_{ig}-\boldsymbol{m}_{ig,0})
   +\begin{pmatrix}\boldsymbol{0}\\
                    \boldsymbol{Y}_i-\boldsymbol{\lambda}_{ig}
     \end{pmatrix},\nonumber\\
 \boldsymbol{H}_{ig}
 &=-\boldsymbol{Q}_{ig}-\boldsymbol{D}_{ig}.
 \label{eq:supp-newton-derivatives}
\end{align}
A Newton proposal is obtained by solving
\begin{equation}
 (\boldsymbol{Q}_{ig}+\boldsymbol{D}_{ig})
 \boldsymbol{\delta}_{ig}=\boldsymbol{g}_{ig},
 \qquad
 \boldsymbol{m}_{ig}^{\mathrm{new}}
 =\boldsymbol{m}_{ig}+\alpha\boldsymbol{\delta}_{ig},
 \label{eq:supp-mean-update}
\end{equation}
where backtracking chooses \(\alpha\in(0,1]\) so that \(B_{ig}\) does not
decrease.

Holding the mean fixed, the stationary equation for the covariance gives the
fixed-point target
\begin{equation}
 \boldsymbol{C}_{ig}^{*}
 =(\boldsymbol{Q}_{ig}+\boldsymbol{D}_{ig})^{-1}.
 \label{eq:supp-cov-target}
\end{equation}
Because \(\boldsymbol{D}_{ig}\) depends on the diagonal of
\(\boldsymbol{C}_{ig,\eta\eta}\), this equation is iterated. In the
implementation, a damped update
\(
\boldsymbol{C}_{ig}^{\mathrm{new}}
=(1-\alpha)\boldsymbol{C}_{ig}+\alpha\boldsymbol{C}_{ig}^{*}
\)
is accepted only when it preserves positive definiteness and does not decrease
the component lower bound $B_{ig}$. The mean and covariance updates are alternated
within each \((i,g)\) pair in the V-step.

After the joint update, Gaussian conditioning converts
\((\boldsymbol{m}_{ig},\boldsymbol{C}_{ig})\) exactly to the notation in
equation~\eqref{eq:supp-var-family}:
\begin{align}
 \boldsymbol{\mu}_{ig}^{(b)}
 &=\boldsymbol{m}_{ig,b},
 &\boldsymbol{S}_{ig}^{(b)}
 &=\boldsymbol{C}_{ig,bb},\nonumber\\
 \boldsymbol{L}_{ig}
 &=\boldsymbol{C}_{ig,\eta b}
   \boldsymbol{C}_{ig,bb}^{-1},
 &\boldsymbol{A}_{ig}
 &=\boldsymbol{m}_{ig,\eta}
   -\boldsymbol{L}_{ig}\boldsymbol{m}_{ig,b},\nonumber\\
 \boldsymbol{V}_{ig}
 &=\boldsymbol{C}_{ig,\eta\eta}
   -\boldsymbol{C}_{ig,\eta b}
    \boldsymbol{C}_{ig,bb}^{-1}
    \boldsymbol{C}_{ig,b\eta}.
 \label{eq:supp-joint-to-conditional}
\end{align}

\subsection{Posterior class-membership and model-parameter updates}
\label{subsec:supp-em-updates}

\paragraph{Class-membership update.} After the V-step, the variational posterior class-membership probabilities are updated as
\begin{equation}
 \tau_{ig}
 =\frac{\pi_{ig}\exp(B_{ig})}
 {\sum_{h=1}^{G}\pi_{ih}\exp(B_{ih})},
 \qquad g=1,\ldots,G,
 \label{eq:supp-tau-update}
\end{equation}
evaluated by a log-sum-exp calculation for numerical stability. Here, $B_{ig}$ is the component lower bound computed in \eqref{eq:supp-component-elbo}.

\paragraph{Model-parameter update.} Let \(N_g=\sum_{i=1}^n\tau_{ig}\). To update the fixed effects, define the
offset-adjusted pseudo-response
\begin{equation*}
 \overline{\boldsymbol{\eta}}_{ig}^{\,*}
 =\boldsymbol{A}_{ig}-\boldsymbol{o}_i
  +(\boldsymbol{L}_{ig}-\boldsymbol{M}_i)
    \boldsymbol{\mu}_{ig}^{(b)}
 =\mathbb{E}_{q_{ig}}(\boldsymbol{\eta}_{ig}
   -\boldsymbol{o}_i-\boldsymbol{M}_i\boldsymbol{b}_{ig}).
\end{equation*}
Using the current \(\boldsymbol{\Sigma}_g\), the generalized least-squares
update is
\begin{equation}
 \boldsymbol{\beta}_g^{\mathrm{new}}
 =\left\{\sum_{i=1}^{n}\tau_{ig}
 \boldsymbol{X}_i^{\trans}\boldsymbol{K}_{\Sigma,ig}^{-1}
 \boldsymbol{X}_i\right\}^{-1}
 \left\{\sum_{i=1}^{n}\tau_{ig}
 \boldsymbol{X}_i^{\trans}\boldsymbol{K}_{\Sigma,ig}^{-1}
 \overline{\boldsymbol{\eta}}_{ig}^{\,*}\right\}.
 \label{eq:supp-beta-update}
\end{equation}

Partition \(\boldsymbol{\mu}_{ig}^{(b)}\) into the outcome-specific subvectors
\(\boldsymbol{\mu}_{ijg}^{(b)}\), and let
\(\boldsymbol{S}_{ig,jj}^{(b)}\) be the \(j\)th diagonal outcome block of
\(\boldsymbol{S}_{ig}^{(b)}\). The constrained update for the random-effect
covariance is
\begin{equation}
 \boldsymbol{\Omega}_g^{(j),\mathrm{new}}
 =\frac{1}{N_g}\sum_{i=1}^{n}\tau_{ig}
 \left\{\boldsymbol{S}_{ig,jj}^{(b)}
 +\boldsymbol{\mu}_{ijg}^{(b)}
  \boldsymbol{\mu}_{ijg}^{(b)\trans}\right\},
 \quad j=1,\ldots,p,
 \label{eq:supp-omega-update}
\end{equation}
followed by
\(\boldsymbol{\Omega}_g^{\mathrm{new}}
=\blockdiag\{\boldsymbol{\Omega}_g^{(1),\mathrm{new}},\ldots,
\boldsymbol{\Omega}_g^{(p),\mathrm{new}}\}\).

For the \(\boldsymbol{\Sigma}_g\) update, recompute
\(\boldsymbol{d}_{ig}\) and \(\boldsymbol{\Psi}_{ig}\) in
equation~\eqref{eq:supp-Psi} using
\(\boldsymbol{\beta}_g^{\mathrm{new}}\). Partition
\(\boldsymbol{\Psi}_{ig}\) into \(p\times p\) visit blocks, and write
\(\boldsymbol{\Psi}_{ig,ll}\) for its \(l\)th diagonal block. Then
\begin{equation}
 \boldsymbol{\Sigma}_g^{\mathrm{new}}
 =\frac{\displaystyle
       \sum_{i=1}^{n}\tau_{ig}\sum_{l=1}^{T_i}
       \boldsymbol{\Psi}_{ig,ll}}
      {\displaystyle
       \sum_{i=1}^{n}\tau_{ig}T_i}.
 \label{eq:supp-sigma-update}
\end{equation}
Only diagonal visit blocks enter this update because the visit-level Gaussian
errors are independent across visits under the model.

Finally, the gating coefficients are updated by maximizing
\begin{equation}
 Q_{\boldsymbol{\xi}}
 :=\sum_{i=1}^{n}\sum_{g=1}^{G}
  \tau_{ig}\log\pi_{ig},
 \label{eq:supp-gating-objective}
\end{equation}
which is a multinomial logistic regression with soft class labels and class
\(G\) as the reference. In our implementation, a weight-decay penalized version of \eqref{eq:supp-gating-objective} is optimized using the \texttt{multinom} function in the \texttt{nnet} package \citep{ripley2002modern, nnet}. The small penalty stabilizes estimation in the presence of near separation. When no penalty is applied and the gating network contains only an intercept, the update has the closed form \(\pi_g^{\mathrm{new}}=N_g/n\) as the mixing proportions.

\subsection{Single-class specialization and overall algorithm}
\label{subsec:supp-g1-algorithm}

For \(G=1\), set \(\pi_{i1}=\tau_{i1} \equiv 1\), remove the gating model and drop the class index \(g\). Equations
\eqref{eq:supp-prior-mean}--\eqref{eq:supp-joint-to-conditional} give the
single-class V-step, whereas equations~\eqref{eq:supp-beta-update},
\eqref{eq:supp-omega-update}, and~\eqref{eq:supp-sigma-update} give its
M-step with \(N_1=n\). Thus, the same updates and the same offset treatment
apply to the single-class and mixture estimators.

For a fixed \(G\), the complete VEM procedure is:
\begin{enumerate}[leftmargin=*,label=\textbf{\arabic*.}]
 \item Initialize the population parameters, gating probabilities when
       \(G>1\), and the joint Gaussian variational moments as described in
       Section~\ref{subsec:supp-initialization}.
 \item For each subject \(i\) and class \(g\), alternate the Newton mean update
       and fixed-point covariance update by following \eqref{eq:supp-mean-update}--\eqref{eq:supp-joint-to-conditional}, and evaluate \(B_{ig}\).
 \item If \(G>1\), update \(\tau_{ig}\) using
       equation~\eqref{eq:supp-tau-update}.
 \item For each $g$, update \(\boldsymbol{\beta}_g\), \(\boldsymbol{\Omega}_g\), and
       \(\boldsymbol{\Sigma}_g\) using
       equations~\eqref{eq:supp-beta-update}--\eqref{eq:supp-sigma-update}.
       If \(G>1\), also update the gating coefficients $\boldsymbol{\xi}$ by maximizing \eqref{eq:supp-gating-objective}.
 \item Re-evaluate the full ELBO in equation~\eqref{eq:supp-mixture-elbo} and
       repeat Steps 2--4 until convergence.
\end{enumerate}

\subsection{Initialization}
\label{subsec:supp-initialization}

\subsubsection{Single-class model}
\label{subsubsec:supp-init-g1}

Initialization is performed on an offset-adjusted approximate log-scale. Define, elementwise,
\begin{equation*}
 \widetilde{\boldsymbol{Y}}_i
 =\log(\boldsymbol{Y}_i+0.5)-\boldsymbol{o}_i.
\end{equation*}
The pseudocount \(0.5\) avoids taking the logarithm of zero. A global ridge least-squares estimate
initializes the fixed effects:
\begin{equation}
 \boldsymbol{\beta}^{(0)}
 =\left\{\sum_{i=1}^{n}\boldsymbol{X}_i^{\trans}\boldsymbol{X}_i
  +\lambda_{\boldsymbol{\beta}}\boldsymbol{I}_{pq}\right\}^{-1}
  \sum_{i=1}^{n}\boldsymbol{X}_i^{\trans}
  \widetilde{\boldsymbol{Y}}_i.
 \label{eq:supp-beta-initial}
\end{equation}
Given \(\boldsymbol{\beta}^{(0)}\), crude subject-specific random effects are
obtained from
\begin{equation*}
 \widehat{\boldsymbol{b}}_i^{(0)}
 =\left(\boldsymbol{M}_i^{\trans}\boldsymbol{M}_i
       +\lambda_{\boldsymbol{b}}\boldsymbol{I}_{pr}\right)^{-1}
   \boldsymbol{M}_i^{\trans}
   \left(\widetilde{\boldsymbol{Y}}_i
        -\boldsymbol{X}_i\boldsymbol{\beta}^{(0)}\right).
\end{equation*}
Each \(\boldsymbol{\Omega}^{(j),(0)}\) is initialized as the empirical covariance matrix of the
\(j\)th outcome-specific blocks of the crude random-effect estimates. To avoid singular or ill-conditioned initial covariance matrices, we add a small diagonal ridge adjustment:
\begin{equation*}
    \boldsymbol{\Omega}^{(j),(0)} = \text{Cov} \left\{\widehat{\boldsymbol{b}}_{ij}^{(0)}: i = 1,\ldots,n \right\} + \lambda_{\boldsymbol{\Omega}} \boldsymbol{I}_r,
\end{equation*}
where $\lambda_{\boldsymbol{\Omega}} > 0$ is a small stabilization constant. 
Next, with
\[
 \widehat{\boldsymbol{\epsilon}}_i^{(0)}
 =\widetilde{\boldsymbol{Y}}_i
  -\boldsymbol{X}_i\boldsymbol{\beta}^{(0)}
  -\boldsymbol{M}_i\widehat{\boldsymbol{b}}_i^{(0)},
\]
\(\boldsymbol{\Sigma}^{(0)}\) is initialized as the empirical covariance matrix of the \(p\)-dimensional visit-level residual vectors, with a small diagonal ridge adjustment:
\begin{equation*}
    \boldsymbol{\Sigma}^{(0)} = \text{Cov}\left\{\widehat{\boldsymbol{\epsilon}}_{il}^{(0)}: i=1,\ldots,n, \, l = 1,\ldots,T_i \right\} + \lambda_{\boldsymbol{\Sigma}}\boldsymbol{I}_{p}.
\end{equation*}
Furthermore, the \(\boldsymbol{\Omega}^{(j),(0)}\) blocks and
\(\boldsymbol{\Sigma}^{(0)}\) are eigenvalue-floored when necessary. This means that if they have very small eigenvalues (nearly singular), we choose a small threshold $\delta > 0$ and replace the eigenvalues smaller than $\delta$ with $\delta$.

For the initial variational state, construct
\(\boldsymbol{Q}_i^{(0)}\) using equation~\eqref{eq:supp-prior-precision} and
the initial population parameters, and set
\begin{align}
 \boldsymbol{m}_i^{(0)}
 &=\begin{pmatrix}\boldsymbol{0}\\
   \boldsymbol{X}_i\boldsymbol{\beta}^{(0)}+\boldsymbol{o}_i
   \end{pmatrix},\nonumber\\
 \boldsymbol{\lambda}_i^{(0)}
 &=\exp\{\boldsymbol{X}_i\boldsymbol{\beta}^{(0)}+
          \boldsymbol{o}_i\},\nonumber\\
 \boldsymbol{C}_i^{(0)}
 &=\left[\boldsymbol{Q}_i^{(0)}+
   \blockdiag\{\boldsymbol{0}_{pr\times pr},
   \diag(\boldsymbol{\lambda}_i^{(0)})\}\right]^{-1}.
 \label{eq:supp-variational-initial}
\end{align}
This uses the model-implied prior precision plus a local Poisson-curvature
approximation.

\subsubsection{Mixture model}
\label{subsubsec:supp-init-mixture}

For \(G>1\), initial labels are obtained from subject-level trajectory
features. For each outcome, these comprise the mean, standard deviation, and
least-squares time slope of
\(
\log(Y_{ijl}+0.5)-o_{il}
\)
over the subject's retained visits. Time-independent gating covariates
\(\boldsymbol{c}_i\), when present, are appended to the trajectory features.
All features are standardized, and \(K\)-means with \(G\) clusters supplies
the initial labels. If an empty cluster occurs, one subject is reassigned from the 
largest cluster containing at least two subjects to the empty cluster. This procedure is repeated
until all $G$ clusters are nonempty.

Within each initial cluster, \(\boldsymbol{\beta}_g^{(0)}\) is obtained by the
ridge regression in equation~\eqref{eq:supp-beta-initial}, restricted to the
subjects assigned to class \(g\). If a cluster is too small to provide stable class-specific
regression estimates, the global fixed-effect estimate is used for that class.
Moreover, a small random perturbation is added to separate similar
class-specific starting values.
The initial \(\boldsymbol{\Omega}_g^{(0)}\) and
\(\boldsymbol{\Sigma}_g^{(0)}\) are then constructed from cluster-specific
crude random effects and visit-level residuals as in the single-class case.
If a cluster contains too few subjects or visits to estimate either covariance,
a positive diagonal matrix is used as a fallback. When \(G>1\), the gating coefficients are initialized by fitting the Softmax model to the initial hard class labels. 

For each \((i,g)\), the initial joint Gaussian mean and covariance are
constructed using \eqref{eq:supp-variational-initial} and the
corresponding class-specific initial parameters. At every subsequent
outer VEM iteration, the joint variational mean
\(\boldsymbol{m}_{ig}\) and covariance \(\boldsymbol{C}_{ig}\) obtained
at the end of the preceding iteration are used as the starting values
for the current class-specific V-step. After the first class-specific
V-step, the initial posterior class-membership probabilities are
computed using \eqref{eq:supp-tau-update}. Multiple initial
clusterings are fitted, and the converged solution with the largest
ELBO is retained.

\subsection{Numerical stabilization and convergence}
\label{subsec:supp-numerics}

The following safeguards are used in implementation:
\begin{itemize}[leftmargin=*]
 \item The Newton and generalized least squares systems are solved using linear-system solvers. Covariance and precision calculations involving symmetric positive-definite matrices are performed using stabilized factorization and inversion routines.
 
 \item Covariance matrices are symmetrized and small eigenvalues are floored. For $\boldsymbol{\Omega}_g$, eigenvalue flooring is applied separately to each outcome-specific diagonal block so that its block-diagonal structure is preserved.
 
 \item The Newton update of the joint variational mean uses a backtracking line search. The step size is initialized at one and successively reduced until the proposed update does not decrease the component lower bound \eqref{eq:supp-component-elbo}. For the joint variational covariance, a damped fixed-point update is used: the current covariance is replaced by a convex combination of its current value and the fixed-point target, with the combination weight reduced when necessary to prevent a decrease in the component lower bound.

 \item If the Newton system is ill-conditioned, the full Hessian is replaced by a diagonal approximation, with an additional diagonal perturbation when necessary. 
 
 \item Exponential arguments are bounded when evaluating the Poisson moments to prevent numerical overflow. The posterior class-membership probabilities are computed using a numerically stable Softmax calculation: for each subject, the largest value of $\log(\pi_{ig})+B_{ig}$ is subtracted from all class-specific values before exponentiation. This is equivalent to the log-sum-exp transformation and does not alter the resulting probabilities.
 
 \item Warm start: at each outer VEM iteration after the first, the joint variational mean $\boldsymbol{m}_{ig}$ and covariance matrix $\boldsymbol{C}_{ig}$ obtained at the end of the preceding outer iteration are used as the initial values for the current class-specific V-step. Similarly, the gating-network optimization is initialized using the gating coefficient estimates from the preceding outer iteration.
\end{itemize}
\paragraph{Convergence criteria.} At outer iteration \(s\), the full ELBO is evaluated after updating the model parameters. The algorithm is declared converged when the absolute change in the ELBO between two successive outer iterations satisfies
\begin{equation*}
 |\mathcal{L}^{(s)}-\mathcal{L}^{(s-1)}| < 10^{-4},
\end{equation*}
or when the maximum number of outer iterations is reached.

\section{Simulation designs and assessment criteria}
\label{sec:supp-sim-design}

We conducted three simulation experiments under the proposed model. Each
configuration was replicated independently \(B=100\) times. The single-class
experiment (\(G=1\)) assessed parameter estimation, the two-class experiment
(\(G=2\)) assessed both parameter estimation and latent-class recovery, and the
three-class experiment (\(G=3\)) assessed latent-class recovery under weak
trajectory separation. Selection of the number of classes was examined for
the \(G=2\) and \(G=3\) experiments. Table~\ref{tab:supp-sim-overview}
summarizes the three designs.

\begin{table}[!ht]
\centering
\small
\caption{Overview of the simulation designs. The dimensions \(q\) and \(r\)
refer to the outcome-specific fixed- and random-effect design vectors,
respectively, and \(q_c\) is the number of gating covariates excluding the
intercept.}
\label{tab:supp-sim-overview}
\begin{tabular}{@{}cccccl@{}}
\toprule
True \(G\) & \(p\) & \((q,r,q_c)\) & \(n\) & Possible \(T_i\) &
Primary assessment \\
\midrule
1 & 3 & \((3,2,0)\) & \(100,300,500\) & \(4,5,6\) &
Parameter estimation \\
2 & 3 & \((3,2,2)\) & \(300,500,700\) & \(3,4,5,6,7\) &
Parameter estimation and ARI \\
3 & 2 & \((4,2,1)\) & \(500\) & \(2,3,4,5\) &
ARI \\
\bottomrule
\end{tabular}
\end{table}

\subsection{Common data-generating mechanism}
\label{subsec:supp-sim-common}

Visit times were generated on \([0,10]\). For each subject, \(T_i\) was drawn
from the setting-specific integer set in Table~\ref{tab:supp-sim-overview},
and \(T_i\) independent uniform times were sorted. The complete visit schedule
was redrawn until every pair of consecutive visits was separated by at least
\(0.5\) time units. We then used the centered and scaled time
\[
 s_{il}=\frac{t_{il}-5}{5},
\]
and generated a time-varying covariate independently across visits as
\[
 x_{il}\mid t_{il}\sim N(0.2s_{il},0.1^2).
\]
For the \(G=1\) and \(G=2\) experiments, the outcome-specific designs were
\[
 \boldsymbol{X}_{ij}(t_{il})=(1,s_{il},x_{il})^{\trans},
 \qquad
 \boldsymbol{R}_{ij}(t_{il})=(1,s_{il})^{\trans}.
\]
For the \(G=3\) experiment, a time-invariant covariate
\(z_i\sim N(0,1)\) was generated independently across subjects and used in
both the longitudinal and gating models, so the outcome-specific designs were
\[
 \boldsymbol{X}_{ij}(t_{il})=(1,s_{il},x_{il},z_i)^{\trans},
 \qquad
 \boldsymbol{R}_{ij}(t_{il})=(1,s_{il})^{\trans}.
\]

For \(G>1\), latent-class membership \(Z_i\) was sampled from the Softmax probabilities in the main
manuscript, with class \(G\) as the reference class. In the \(G=2\) experiment,
the gating covariates were
\(\boldsymbol{c}_i=(c_{i1},c_{i2})^{\trans}\), where
\(c_{i1},c_{i2}\stackrel{\mathrm{iid}}{\sim}N(0,1)\). In the \(G=3\)
experiment, \(\boldsymbol{c}_i=z_i\), where $z_i$ is the aforementioned time-invariant covariate used in the fixed design. Accordingly, class sizes were random
and were not fixed in advance.

Conditional on \(Z_i=g\), the random effects and visit-level Gaussian errors
were generated as
\[
 \boldsymbol{b}_{ig}\sim
 N(\boldsymbol{0},\boldsymbol{\Omega}_g),
 \qquad
 \boldsymbol{\epsilon}_{ilg}\sim
 N(\boldsymbol{0},\boldsymbol{\Sigma}_g),
\]
independently across subjects and, for the visit-level errors, independently
across visits. Counts were then generated from the Poisson-lognormal mixed model. The known offsets were set to \(o_{il}=0\) in every simulation.
All \(p\) outcomes were observed at every generated visit, so no visits were
deleted after generation. Irregularity arose from the subject-specific numbers
and timings of visits.

\subsection{Parameter settings}
\label{subsec:supp-sim-parameters}

For compact presentation, the fixed-effect values below are arranged in
\(p\times q\) matrices whose \(j\)th row is
\(\boldsymbol{\beta}_{gj}^{\trans}\): 
\[
 \boldsymbol{\beta}_g
 =
 \left(
 \boldsymbol{\beta}_{g1}^{\trans},\ldots,
 \boldsymbol{\beta}_{gp}^{\trans}
 \right)^{\trans}
\]
For \(G=1\), the class
subscript $g$ is omitted. In every setting,
\[
 \boldsymbol{\Omega}_g
 =
 \blockdiag\left\{
 \boldsymbol{\Omega}^{(1)}_g,\ldots,
 \boldsymbol{\Omega}^{(p)}_g
 \right\}.
\]

\subsubsection{Single-class setting}

For \(G=1\), the fixed-effect coefficients were
\[
\begin{pmatrix}
\boldsymbol{\beta}_1^{\trans}\\\\
\boldsymbol{\beta}_2^{\trans}\\\\
\boldsymbol{\beta}_3^{\trans}
\end{pmatrix}
=
\begin{pmatrix}
 6.00 &  1.00 & -1.50\\
 5.80 & -1.30 &  1.20\\
 7.00 &  1.20 &  0.80
\end{pmatrix}.
\]
The outcome-specific random-effect covariance blocks and the visit-level
covariance matrix were
\[
\begin{aligned}
\boldsymbol{\Omega}^{(1)}
&=
\begin{pmatrix}
1.00&0.30\\
0.30&1.00
\end{pmatrix},
&
\boldsymbol{\Omega}^{(2)}
&=
\begin{pmatrix}
0.80&0.20\\
0.20&0.70
\end{pmatrix},
&
\boldsymbol{\Omega}^{(3)}
&=
\begin{pmatrix}
0.60&-0.25\\
-0.25&1.10
\end{pmatrix},
\end{aligned}
\]
and
\[
\boldsymbol{\Sigma}
=
\begin{pmatrix}
 0.50 &  0.18 & -0.16\\
 0.18 &  0.45 & -0.14\\
-0.16 & -0.14 &  0.40
\end{pmatrix}.
\]

\subsubsection{Two-class setting}

For \(G=2\), the class-specific fixed-effect coefficients were
\[
\begin{pmatrix}
\boldsymbol{\beta}_{11}^{\trans}\\\\
\boldsymbol{\beta}_{12}^{\trans}\\\\
\boldsymbol{\beta}_{13}^{\trans}
\end{pmatrix}
=
\begin{pmatrix}
5.70& 0.25&-0.50\\
5.95& 0.65&-0.50\\
6.55&-0.30& 0.65
\end{pmatrix},
\qquad
\begin{pmatrix}
\boldsymbol{\beta}_{21}^{\trans}\\\\
\boldsymbol{\beta}_{22}^{\trans}\\\\
\boldsymbol{\beta}_{23}^{\trans}
\end{pmatrix}
=
\begin{pmatrix}
5.75&-2.00&1.60\\
5.90& 2.30&1.55\\
6.30&-3.00&1.80
\end{pmatrix}.
\]
The random-effect covariance blocks were
\[
\begin{aligned}
\boldsymbol{\Omega}^{(1)}_1
&=
\begin{pmatrix}
0.80&0.10\\
0.10&0.30
\end{pmatrix},
&
\boldsymbol{\Omega}^{(2)}_1
&=
\begin{pmatrix}
0.65&0.12\\
0.12&0.32
\end{pmatrix},
&
\boldsymbol{\Omega}^{(3)}_1
&=
\begin{pmatrix}
0.75&-0.10\\
-0.10&0.30
\end{pmatrix},
\\[4pt]
\boldsymbol{\Omega}^{(1)}_2
&=
\begin{pmatrix}
0.70&-0.24\\
-0.24&0.55
\end{pmatrix},
&
\boldsymbol{\Omega}^{(2)}_2
&=
\begin{pmatrix}
0.85&0.30\\
0.30&0.65
\end{pmatrix},
&
\boldsymbol{\Omega}^{(3)}_2
&=
\begin{pmatrix}
0.68&0.22\\
0.22&0.58
\end{pmatrix}.
\end{aligned}
\]
The visit-level covariance matrices were
\[
\boldsymbol{\Sigma}_1
=
\begin{pmatrix}
 0.55& 0.18&-0.16\\
 0.18& 0.50&-0.14\\
-0.16&-0.14& 0.48
\end{pmatrix},
\qquad
\boldsymbol{\Sigma}_2
=
\begin{pmatrix}
 0.58&-0.15&0.16\\
-0.15& 0.55&0.18\\
 0.16& 0.18&0.54
\end{pmatrix}.
\]
With Class~2 as the reference class, the gating coefficient for Class~1 was
\[
 \boldsymbol{\xi}_1=(0.40,1.00,-1.00)^{\trans}.
\]

\subsubsection{Three-class setting}

For \(G=3\), the class-specific fixed-effect coefficients were
\[
\begin{aligned}
\begin{pmatrix}
\boldsymbol{\beta}_{11}^{\trans}\\\\
\boldsymbol{\beta}_{12}^{\trans}
\end{pmatrix}
&=
\begin{pmatrix}
5.80&1.10&-1.25& 0.65\\
6.45&2.85&-1.20& 0.90
\end{pmatrix},
&
\begin{pmatrix}
\boldsymbol{\beta}_{21}^{\trans}\\\\
\boldsymbol{\beta}_{22}^{\trans}
\end{pmatrix}
&=
\begin{pmatrix}
5.90& 1.00&-1.15& 0.60\\
5.75&-1.85& 1.25&-0.90
\end{pmatrix},
\\[4pt]
\begin{pmatrix}
\boldsymbol{\beta}_{31}^{\trans}\\\\
\boldsymbol{\beta}_{32}^{\trans}
\end{pmatrix}
&=
\begin{pmatrix}
5.75&1.05&-1.30&0.70\\
6.15&0.25& 1.85&0.60
\end{pmatrix}.&&
\end{aligned}
\]
The random-effect covariance blocks were
\[
\begin{aligned}
\boldsymbol{\Omega}^{(1)}_1
&=
\begin{pmatrix}
0.30&0.05\\
0.05&0.14
\end{pmatrix},
&
\boldsymbol{\Omega}^{(2)}_1
&=
\begin{pmatrix}
0.34&0.07\\
0.07&0.18
\end{pmatrix},
\\[4pt]
\boldsymbol{\Omega}^{(1)}_2
&=
\begin{pmatrix}
0.32&0.04\\
0.04&0.14
\end{pmatrix},
&
\boldsymbol{\Omega}^{(2)}_2
&=
\begin{pmatrix}
0.34&-0.07\\
-0.07&0.18
\end{pmatrix},
\\[4pt]
\boldsymbol{\Omega}^{(1)}_3
&=
\begin{pmatrix}
0.30&0.04\\
0.04&0.13
\end{pmatrix},
&
\boldsymbol{\Omega}^{(2)}_3
&=
\begin{pmatrix}
0.32&0.06\\
0.06&0.17
\end{pmatrix}.
\end{aligned}
\]
The visit-level covariance matrices were
\[
\boldsymbol{\Sigma}_1
=
\begin{pmatrix}
0.15&0.04\\
0.04&0.13
\end{pmatrix},
\quad
\boldsymbol{\Sigma}_2
=
\begin{pmatrix}
0.15&0.035\\
0.035&0.13
\end{pmatrix},
\quad
\boldsymbol{\Sigma}_3
=
\begin{pmatrix}
0.16&0.04\\
0.04&0.14
\end{pmatrix}.
\]
With Class~3 as the reference class, the gating coefficients were
\[
 \boldsymbol{\xi}_1=(-0.45,1.25)^{\trans},
 \qquad
 \boldsymbol{\xi}_2=(-0.45,-1.25)^{\trans}.
\]

\subsection{Fitting the proposed model and selecting the number of classes}
\label{subsec:supp-sim-fitting}

All proposed models were fitted in a completely unsupervised fashion. The data-adaptive initialization described in
Section~\ref{subsec:supp-initialization} was used. The correctly specified
\(G=1\) and \(G=2\) models were fitted from one such initialization. Because
the \(G=3\) setting was more susceptible to local optima, it was fitted from 20 independently seeded \(k\)-means-based
initializations described in Section~\ref{subsec:supp-initialization}, and the fit with the largest final ELBO was retained. The remaining numerical safeguards and stopping rules were those described in
Section~\ref{subsec:supp-numerics}.

For data generated with \(G=2\), candidate models with
\(G\in\{1,2,3\}\) were compared. For data generated with \(G=3\), candidate
models with \(G\in\{1,2,3,4\}\) were compared. For each fitted model
\(\mathcal{M}_G\), we computed its variational BIC
\[
 \widetilde{\operatorname{BIC}}(\mathcal{M}_G)
 =
 -2\mathcal{L}_{\mathcal{M}_G}
 \left(
 \widehat q_{\mathcal{M}_G},
 \widehat{\boldsymbol{\theta}}_{\mathcal{M}_G}
 \right)
 +\nu_G\log n,
\]
and selected the candidate model with the smallest variational BIC. With \(q_c\) gating
covariates and class \(G\) as the reference, the number of free parameters was
\[
 \nu_G
 =
 G\left\{
 pq+
 p\frac{r(r+1)}{2}+
 \frac{p(p+1)}{2}
 \right\}
 +(G-1)(q_c+1).
\]
The resulting counts were \(24,51,\) and \(78\) for candidate
\(G=1,2,3\), respectively, under the two-class design, and \(17,36,55,\)
and \(74\) for candidate \(G=1,2,3,4\), respectively, under the three-class
design. Subject-specific variational parameters were not included in
\(\nu_G\), and the known offset contributed no free parameter.

\subsection{Comparator configurations}
\label{subsec:supp-sim-comparators}

Parameter estimation was compared with \texttt{lme4} only for \(G=1\),
whereas \texttt{lcmm} and \texttt{clustGLMM} were compared with the proposed
method only in terms of latent-class recovery and selection of \(G\).

\subsubsection{
  \texorpdfstring{\texttt{lme4} benchmark for \(G=1\)}
  {lme4 benchmark for G = 1}
}

The single-class datasets with $n = 100, 300, 500$ were fitted using
\texttt{lme4::glmer} respectively with a Poisson response and log link
\citep{bates2015lme4}. The three outcomes were stacked in long format. The
model included outcome-specific intercept, scaled-time, and time-varying
covariate effects; separate outcome-specific subject random intercepts and
random scaled-time slopes, each with an unrestricted \(2\times2\) covariance
matrix; and a visit-specific three-dimensional random intercept with an
unrestricted covariance matrix. The latter term provides the
\texttt{lme4} analogue of the visit-level covariance
\(\boldsymbol{\Sigma}\). The Laplace approximation was used
(\texttt{nAGQ = 1}); optimizer used \texttt{bobyqa} followed, when
needed, by \texttt{Nelder\_Mead}, with a maximum of \(2\times10^5\)
function evaluations.

\subsubsection{
  \texorpdfstring{\texttt{lcmm} benchmarks for \(G=2\) and \(G=3\)}
  {lcmm benchmarks for G = 2 and G = 3}
}

Because \texttt{lcmm} does not provide the multivariate Poisson-lognormal
model used to generate the data, each count outcome was transformed to
\(\log(Y_{ijl}+0.5)\) and fitted as a Gaussian longitudinal process
\citep{proust2017estimation}. Outcome-specific \texttt{hlme} submodels were
combined using \texttt{mpjlcmm}, with a shared latent-class variable and
class-membership model. In the \(G=2\) experiment, each outcome model included
an intercept, scaled time, and the time-varying covariate $x_{il}$; all three effects were allowed to
vary by class, and the class-membership model included \(c_{i1}\) and
\(c_{i2}\). In the \(G=3\) experiment, \(z_i\) was additionally included in
each outcome model and was used as the class-membership covariate. Each
outcome had a correlated random intercept and random scaled-time slope
(\texttt{idiag = FALSE}). Setting \texttt{nwg = TRUE} allowed proportional
class-specific random-effect covariance matrices, the closest available
\texttt{lcmm} specification to the data-generating model.

One-class fits supplied the initial models for the \texttt{lcmm} grid search.
For the correctly specified \(G=2\) and \(G=3\) fits, respectively, one and
five short random-start runs of at most 50 iterations were used; the best run
was then optimized for at most 500 iterations. The candidate grids for model
selection matched those used for the proposed method, and the candidate with
the smallest \texttt{lcmm} BIC was selected. Unlike the proposed model, this
benchmark used transformed Gaussian outcomes and did not contain a direct analogue of the unrestricted class-specific visit-level covariance matrix
\(\boldsymbol{\Sigma}_g\).

\subsubsection{
  \texorpdfstring{\texttt{clustGLMM} benchmarks for \(G=2\) and \(G=3\)}
  {clustGLMM benchmarks for G = 2 and G = 3}
}

The \texttt{clustGLMM} benchmark used the original counts with Poisson log-link
models \citep{vavra2024clusterwise}. The class-specific covariate effects and
random intercept--slope designs matched the corresponding longitudinal
designs above. However, \texttt{clustGLMM} used constant mixing weights and
did not include the gating covariates or a separate visit-level multivariate
Gaussian error. Thus, direct comparison of its parameter estimates with the
generating parameters was not appropriate.

For each dataset, both a fixed-\(G\) fit and a sparse finite-mixture fit were
run. The fixed-\(G\) fit used the true number of classes. The sparse fit used
\(G_{\max}=3\) for the two-class experiment and \(G_{\max}=4\) for the
three-class experiment. Two MCMC chains were run for 15,000 iterations each;
the first 5,000 iterations were discarded as burn-in, leaving 10,000
post-burn-in iterations per chain. Every fifth post-burn-in draw was retained
for posterior clustering summaries. Regression coefficients were assigned
\(N(0,5^2)\) priors, the stored scaled time was used without further
standardization, and trajectory-based initial values were constructed from
ridge regressions of \(\log(Y+0.5)\) followed by \(k\)-means. The transformed
counts were used only to initialize the sampler; the fitted likelihood used
the original Poisson counts.

For the sparse fit, the selected number of classes was the posterior mode of
the number \(G^+\) of occupied components over the pooled post-burn-in draws;
no BIC was computed. Subject classifications were the maximum-posterior
assignments obtained from the relabelled allocation frequencies conditional
on the selected \(G^+\). The primary model-selection comparison used these
sparse-fit partitions.

\subsection{Assessment criteria}
\label{subsec:supp-sim-assessment}

\subsubsection{Parameter estimation}

For replication \(b\), parameter recovery was measured by the relative
\(L_2\) error for the stacked fixed-effect vector and the relative Frobenius
error for each covariance matrix:
\[
 \operatorname{RE}_{\boldsymbol{\beta}_g}^{(b)}
 =
 \frac{
 \left\|
 \widehat{\boldsymbol{\beta}}_g^{(b)}
 -
 \boldsymbol{\beta}_g
 \right\|_2
 }{
 \left\|
 \boldsymbol{\beta}_g
 \right\|_2
 },
 \qquad
 \operatorname{RE}_{\boldsymbol{A}_g}^{(b)}
 =
 \frac{
 \left\|
 \widehat{\boldsymbol{A}}_g^{(b)}
 -
 \boldsymbol{A}_g
 \right\|_{\mathrm F}
 }{
 \left\|
 \boldsymbol{A}_g
 \right\|_{\mathrm F}
 },
 \quad
 \boldsymbol{A}_g\in
 \left\{
 \boldsymbol{\Omega}_g,
 \boldsymbol{\Sigma}_g
 \right\}.
\]
For \(\boldsymbol{\Omega}_g\), the Frobenius norm was computed using only
the structurally free entries in its outcome-specific diagonal blocks. For every parameter block and sample
size, we report the mean and standard deviation of the replication-level
relative errors. These are the quantities displayed in the parameter-recovery
figure in the main manuscript and tabulated in
Section~\ref{sec:supp-sim-results}.

For \(G=2\), fitted component labels were aligned with the generating labels
separately in each replication before parameter errors were calculated. Let
\(\operatorname{bvec}(\boldsymbol{\Omega})\) stack only the entries in the
free within-outcome blocks. The distance between generating class \(g\) and
fitted class \(h\) was
\[
 d_{g,h}
 =
 \frac{
 \left\|
 \widehat{\boldsymbol{\beta}}_h
 -
 \boldsymbol{\beta}_g
 \right\|_2^2
 }{
 \left\|
 \boldsymbol{\beta}_g
 \right\|_2^2
 }
 +
 \frac{
 \left\|
 \operatorname{bvec}(\widehat{\boldsymbol{\Omega}}_h)
 -
 \operatorname{bvec}(\boldsymbol{\Omega}_g)
 \right\|_2^2
 }{
 \left\|
 \operatorname{bvec}(\boldsymbol{\Omega}_g)
 \right\|_2^2
 }
 +
 \frac{
 \left\|
 \widehat{\boldsymbol{\Sigma}}_h
 -
 \boldsymbol{\Sigma}_g
 \right\|_{\mathrm F}^2
 }{
 \left\|
 \boldsymbol{\Sigma}_g
 \right\|_{\mathrm F}^2
 }.
\]
Of the two possible permutations \(\rho\) of the fitted labels, we selected
the one minimizing \(\sum_{g=1}^2d_{g,\rho(g)}\). 
No parameter-recovery analysis was conducted for \(G=3\), as that experiment was designed to assess
clustering under weak trajectory separation.

\subsubsection{Latent-class recovery}

For the proposed method, subjects were assigned by maximum posterior
class-membership probability,
\[
 \widehat Z_i
 =
 \mathop{\arg\max}_{1\leq g\leq G}
 \widehat\tau_{ig}.
\]
The corresponding maximum-posterior classifications were used for the two
comparators. Agreement with the true partition was measured by the adjusted Rand index (ARI) \citep{HubertArabie1985}. 
The ARI is invariant to class-label permutations, so no label alignment is
required for this measure. We summarize its distribution across replications
using the mean, median, standard deviation, and interquartile range (IQR) in Section~\ref{sec:supp-sim-results}.

\subsubsection{Selection of the number of classes}

For the proposed method and \texttt{lcmm}, model-selection performance was
summarized by the number and percentage of replications in which the information
criterion selected each candidate value of \(G\), using the variational BIC for our proposed approach
and the classical BIC for \texttt{lcmm}, respectively. For \texttt{clustGLMM}, the same
summary was formed from the posterior modal occupied-component count \(G^+\)
from the sparse fit. 

\section{Additional simulation results}
\label{sec:supp-sim-results}

\subsection{Single-class parameter estimation and comparison with
\texttt{lme4}}
\label{subsec:supp-g1-results}

We first examined parameter estimation in the single-class setting with
\(G=1\), \(p=3\), and \(n\in\{100,300,500\}\). For each sample size, the
proposed VEM procedure and the \texttt{lme4} benchmark described in
Section~\ref{subsec:supp-sim-comparators} were fitted to the same 100
independently generated datasets. Both methods estimated the fixed-effect
coefficients \(\boldsymbol{\beta}\), the block-diagonal subject-level
random-effect covariance matrix \(\boldsymbol{\Omega}\), and the visit-level
covariance matrix \(\boldsymbol{\Sigma}\).

Parameter recovery was evaluated using the relative \(L_2\) error for
\(\boldsymbol{\beta}\) and the relative Frobenius errors for
\(\boldsymbol{\Omega}\) and \(\boldsymbol{\Sigma}\), as defined in
Section~\ref{subsec:supp-sim-assessment}. For
\(\boldsymbol{\Omega}\), only the structurally free entries within its three
outcome-specific diagonal blocks contributed to the error. The off-block
entries of both the generating and fitted \texttt{lme4} covariance matrices
were fixed at zero, so computing the Frobenius error from the full
block-diagonal matrix gives the same result.

Table~\ref{tab:supp-g1-relative-errors} reports the mean and standard deviation
of each replication-level relative error. All fits converged at each sample size, with no fitting failures or singular fits.

\begin{table}[!ht]
\centering
\small
\caption{
Relative parameter-estimation errors in the single-class simulation.
Values are the mean (standard deviation) across 100 replications. Minor
differences between the two methods occur beyond the displayed precision.
}
\label{tab:supp-g1-relative-errors}
\begin{tabular}{@{}llccc@{}}
\toprule
Parameter block
& Method
& \(n=100\)
& \(n=300\)
& \(n=500\) \\
\midrule
\(\boldsymbol{\beta}\)
& Proposed VEM
& 0.0602 (0.0216)
& 0.0360 (0.0112)
& 0.0273 (0.0105) \\
\(\boldsymbol{\beta}\)
& \texttt{lme4}
& 0.0602 (0.0216)
& 0.0360 (0.0112)
& 0.0273 (0.0105) \\
\addlinespace

\(\boldsymbol{\Omega}\)
& Proposed VEM
& 0.2118 (0.0548)
& 0.1170 (0.0320)
& 0.0899 (0.0245) \\
\(\boldsymbol{\Omega}\)
& \texttt{lme4}
& 0.2118 (0.0548)
& 0.1170 (0.0320)
& 0.0899 (0.0245) \\
\addlinespace

\(\boldsymbol{\Sigma}\)
& Proposed VEM
& 0.1063 (0.0436)
& 0.0605 (0.0186)
& 0.0465 (0.0144) \\
\(\boldsymbol{\Sigma}\)
& \texttt{lme4}
& 0.1063 (0.0436)
& 0.0605 (0.0186)
& 0.0465 (0.0144) \\
\bottomrule
\end{tabular}
\end{table}

\begin{figure}[!ht]
\centering
\includegraphics[
    width=0.72\textwidth
]{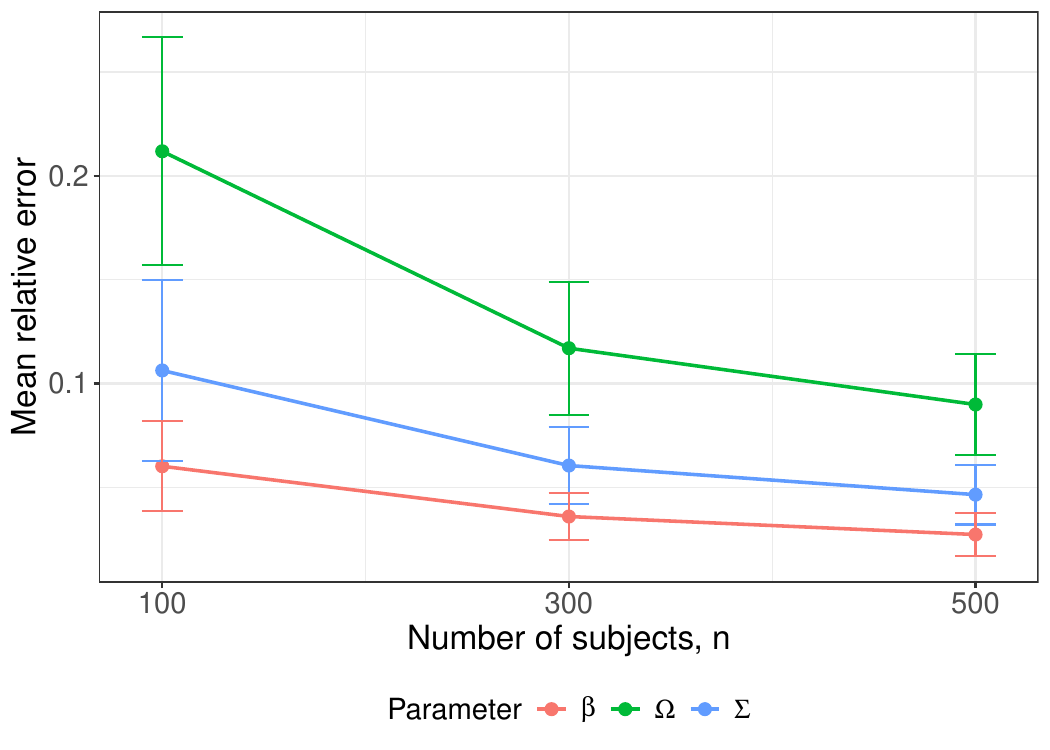}
\caption{
Parameter recovery for the proposed VEM estimator in the single-class
simulation. Points represent the mean relative errors across 100 replications,
and error bars extend one standard deviation above and below the mean. Relative
\(L_2\) error is shown for \(\boldsymbol{\beta}\), whereas relative Frobenius
errors are shown for \(\boldsymbol{\Omega}\) and
\(\boldsymbol{\Sigma}\). For \(\boldsymbol{\Omega}\), the error was calculated
using only its structurally free within-outcome blocks.
}
\label{fig:supp-g1-relative-errors}
\end{figure}

For the proposed VEM estimator, Figure~\ref{fig:supp-g1-relative-errors}
shows how the mean relative errors changed with sample size. 

As displayed in Table~\ref{tab:supp-g1-relative-errors}, estimation accuracy improved consistently with increasing sample size $n$. 
For the proposed method, increasing \(n\) from 100 to 500 reduced the mean relative
errors by approximately \(54.66\%\), \(57.56\%\), and \(56.26\%\) for
\(\boldsymbol{\beta}\), \(\boldsymbol{\Omega}\), and
 \(\boldsymbol{\Sigma}\), respectively. 
The standard deviations also decreased,
indicating greater stability as the number of subjects increased.

Across all sample sizes, the fixed-effect coefficients were estimated most
accurately, followed by the visit-level covariance matrix. The
subject-level random-effect covariance matrix had the largest relative error,
reflecting the greater difficulty of estimating multiple random-intercept and
random-slope variances and covariances from a limited number of repeated
observations per subject. Nevertheless, its mean relative error decreased from \(0.2118\) at \(n=100\) to \(0.0899\) at \(n=500\).

Overall, the proposed VEM and \texttt{lme4} procedures produced nearly identical
relative errors for every parameter block and sample size. This close
agreement provides a numerical validation of the proposed estimation
procedure in the homogeneous setting, where the stacked Poisson generalized
linear mixed model fitted by \texttt{lme4} closely matches the covariance
structure of the generating model. 

\subsection{Numerical summaries underlying the main-text figures}
\label{subsec:supp-main-figure-summaries}

This subsection reports the numerical summaries underlying the
parameter-estimation and latent-class-recovery panels of the main-text
simulation figures. All summaries are based on 100 independent replications.

\paragraph{Parameter recovery for \(G=2\).}
Table~\ref{tab:supp-g2-relative-error} reports the mean and standard
deviation (SD) of the blockwise relative errors. Before the errors were
calculated, the fitted component labels were aligned with the generating
classes within each replication, as described in
Section~\ref{sec:supp-sim-design}. For \(\boldsymbol{\beta}_g\), relative
error was measured using the $L_2$ norm of the vectorized coefficient
block. For \(\boldsymbol{\Omega}_g\) and
\(\boldsymbol{\Sigma}_g\), relative error was measured using the Frobenius
norm. Only the structurally free within-outcome blocks of
\(\boldsymbol{\Omega}_g\) were included in the calculation.

\begin{table}[!htbp]
\centering
\caption{Blockwise relative errors for the proposed method in the
\(G=2\) simulation setting. Each entry is the mean (SD) across 100
independent replications after alignment of the fitted and generating
class labels.}
\label{tab:supp-g2-relative-error}
\small
\setlength{\tabcolsep}{4pt}
\begin{tabular}{@{}c cc cc cc@{}}
\toprule
& \multicolumn{2}{c}{Fixed effects}
& \multicolumn{2}{c}{Random-effect covariance}
& \multicolumn{2}{c}{Visit-level covariance} \\
\cmidrule(lr){2-3}
\cmidrule(lr){4-5}
\cmidrule(lr){6-7}
\(n\)
& \(\boldsymbol{\beta}_1\) & \(\boldsymbol{\beta}_2\)
& \(\boldsymbol{\Omega}_1\) & \(\boldsymbol{\Omega}_2\)
& \(\boldsymbol{\Sigma}_1\) & \(\boldsymbol{\Sigma}_2\) \\
\midrule
300 & 0.0555 (0.0216) & 0.0585 (0.0220)
    & 0.1889 (0.0438) & 0.2189 (0.0608)
    & 0.0858 (0.0305) & 0.1142 (0.0387) \\
500 & 0.0376 (0.0159) & 0.0474 (0.0187)
    & 0.1422 (0.0323) & 0.1756 (0.0409)
    & 0.0668 (0.0250) & 0.0785 (0.0279) \\
700 & 0.0319 (0.0120) & 0.0402 (0.0153)
    & 0.1174 (0.0276) & 0.1378 (0.0356)
    & 0.0562 (0.0183) & 0.0695 (0.0247) \\
\bottomrule
\end{tabular}
\end{table}

As displayed in Table~\ref{tab:supp-g2-relative-error}, the mean relative error decreased with increasing sample size for every parameter block, and the corresponding SDs also decreased. From \(n=300\)
to \(n=700\), the mean relative errors decreased by \(42.63\%\) and
\(31.21\%\) for \(\boldsymbol{\beta}_1\) and
\(\boldsymbol{\beta}_2\), respectively; by \(37.85\%\) and \(37.07\%\)
for \(\boldsymbol{\Omega}_1\) and \(\boldsymbol{\Omega}_2\),
respectively; and by \(34.49\%\) and \(39.10\%\) for
\(\boldsymbol{\Sigma}_1\) and \(\boldsymbol{\Sigma}_2\),
respectively. The random-effect
covariance matrices remained the most difficult blocks to estimate, but
their errors declined substantially as \(n\) increased.

\paragraph{Latent-class recovery.}
For each replication, subjects were assigned to the class with the largest
estimated posterior class-membership probability, and agreement with the
true partition was measured using the ARI discussed in Section~\ref{sec:supp-sim-design}.
Table~\ref{tab:supp-proposed-ari} summarizes the resulting ARI
distributions over 100 replications for each simulation setting. The \(Q_1\)--\(Q_3\) column gives the interquartile
interval, whereas the IQR column gives its width, \(Q_3-Q_1\).

\begin{table}[!htbp]
\centering
\caption{ARI summaries for the proposed method across
100 independent replications.}
\label{tab:supp-proposed-ari}
\small
\setlength{\tabcolsep}{8pt}
\begin{tabular}{@{}ccccccc@{}}
\toprule
True \(G\) & \(n\) & Mean & Median & SD & \(Q_1\)--\(Q_3\) & IQR \\
\midrule
2 & 300 & 0.917 & 0.909 & 0.032 & 0.896--0.947 & 0.051 \\
2 & 500 & 0.924 & 0.929 & 0.023 & 0.906--0.945 & 0.039 \\
2 & 700 & 0.923 & 0.927 & 0.022 & 0.911--0.944 & 0.033 \\
3 & 500 & 0.801 & 0.802 & 0.029 & 0.783--0.821 & 0.038 \\
\bottomrule
\end{tabular}
\end{table}

As displayed in Table~\ref{tab:supp-proposed-ari}, latent-class recovery was strong in all three \(G=2\) settings. The mean
and median ARIs exceeded 0.90, and the reduction in both the SD and IQR
from \(n=300\) to \(n=700\) indicates that recovery became more consistent
as the sample size increased. The mean and median were similar at
\(n=500\) and \(n=700\), suggesting that recovery had largely stabilized
over this sample-size range. On the other hand, the \(G=3\), \(n=500\) setting was more challenging because of the weaker
separation among the class-specific trajectories. Nevertheless, its mean
and median ARIs were both approximately 0.80, with an SD of 0.029 and an
interquartile interval of 0.783--0.821, indicating substantial and stable
agreement with the true partition.

\subsection{Entrywise covariance recovery}
\label{subsec:supp-covariance-recovery}

The blockwise relative errors reported in the preceding subsections
summarize the overall recovery of each covariance matrix but do not show
how estimation accuracy varies across individual variance and covariance
entries. We therefore examined entrywise mean absolute estimation errors.
For
\(\mathbf{A}_g\in
\{\boldsymbol{\Omega}_g,\boldsymbol{\Sigma}_g\}\),
the mean absolute error of entry \((u,v)\) was calculated as
\[
\operatorname{MAE}_{g,uv}(\mathbf{A})
=
\frac{1}{B}\sum_{m=1}^{B}
\left|
\widehat{A}_{g,uv}^{(m)}-A_{g,uv}
\right|,
\qquad B=100,
\]
where $A_{g,uv}$ is the corresponding true parameter value, and \(m\) indexes the simulation replications. For the \(G=2\)
setting, the fitted component labels were aligned with the generating
classes within each replication before the entrywise errors were
calculated. Because \(\boldsymbol{\Omega}_g\) is block diagonal across
outcomes, only its structurally free within-outcome blocks were evaluated.

\begin{figure}[t]
\centering
\textbf{A}\par\vspace{0.2em}
\includegraphics[width=0.98\textwidth]
{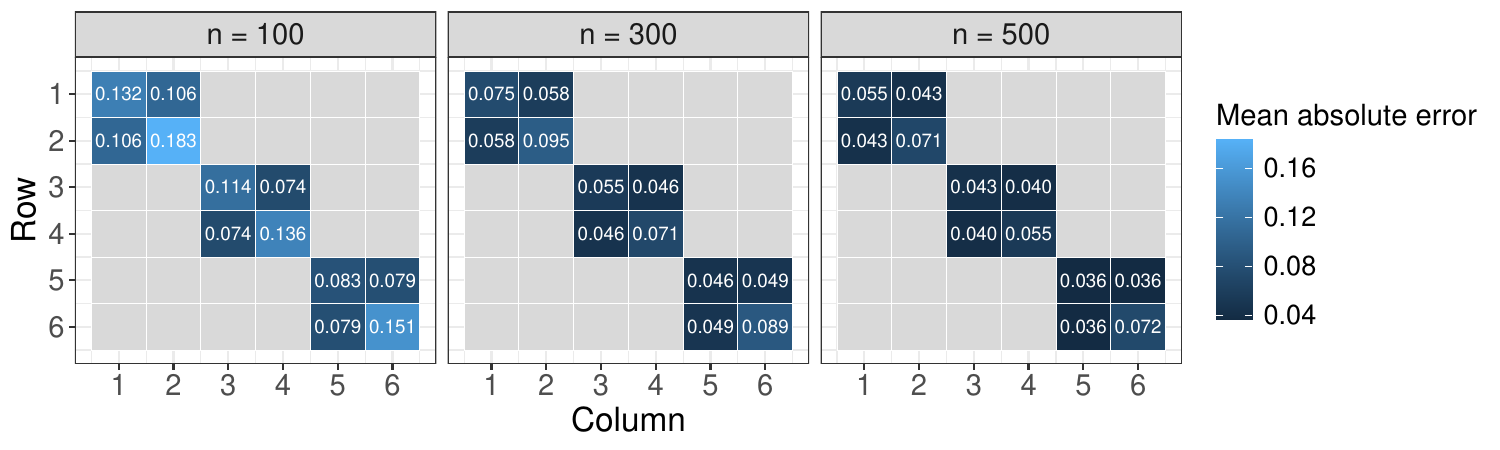}

\vspace{0.8em}
\textbf{B}\par\vspace{0.2em}
\includegraphics[width=0.93\textwidth]
{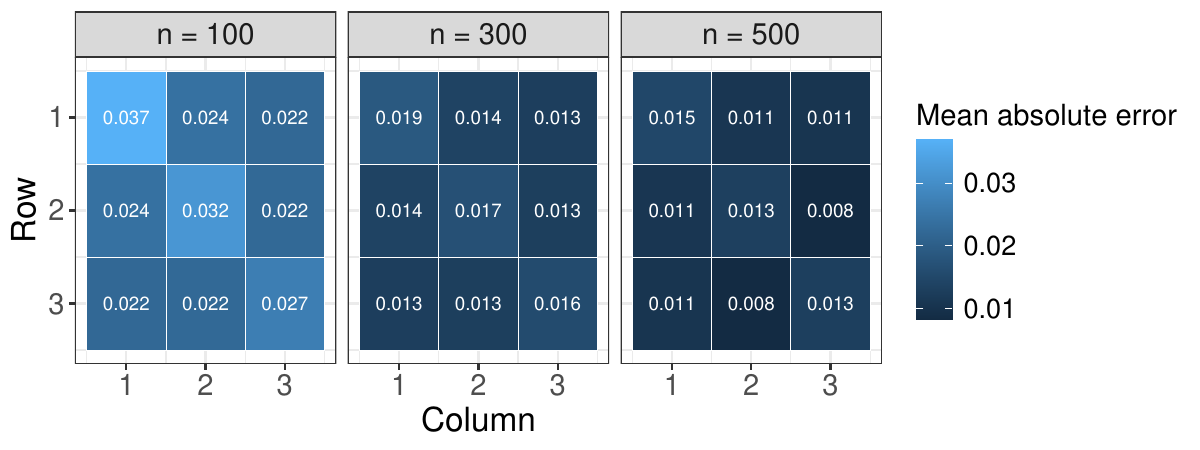}

\caption{
Entrywise covariance recovery in the \(G=1\) simulation setting.
Heatmaps show the mean absolute estimation error across 100 independent
replications for \textbf{(A)} the random-effect covariance matrix
\(\boldsymbol{\Omega}\) and \textbf{(B)} the visit-level covariance
matrix \(\boldsymbol{\Sigma}\), for
\(n\in\{100,300,500\}\). Numerical labels give the entrywise mean
absolute errors. In panel \textbf{(A)}, grey cells represent structural
zeros imposed by the block-diagonal specification of
\(\boldsymbol{\Omega}\) and were not included in the error calculation.
A common color scale is used across sample sizes within each panel.
}
\label{fig:supp-g1-covariance-mae}
\end{figure}

\begin{figure}[H]
\centering
\textbf{A}\par\vspace{0.2em}
\includegraphics[width=0.63\textwidth]
{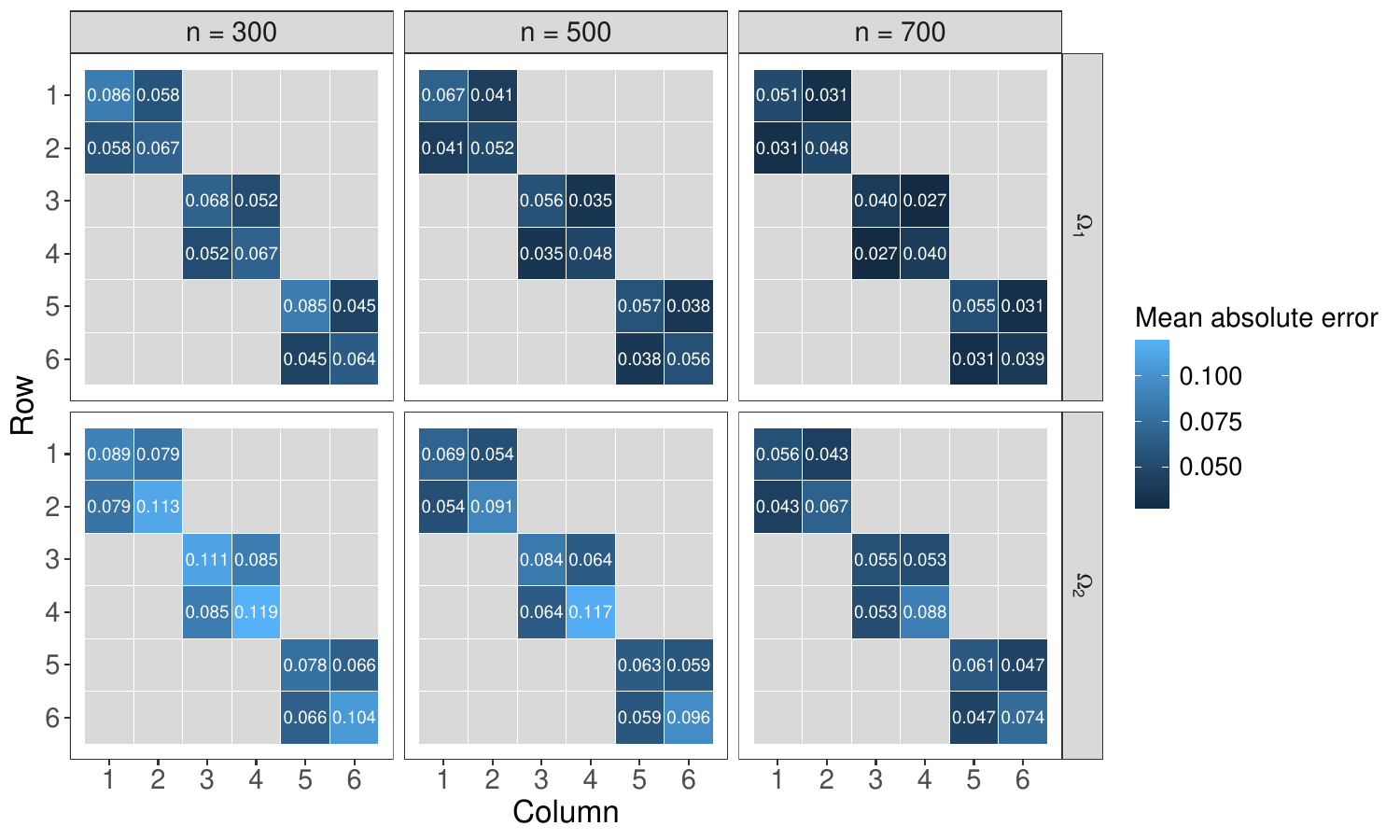}

\vspace{0.4em}
\textbf{B}\par\vspace{0.2em}
\includegraphics[width=0.63\textwidth]
{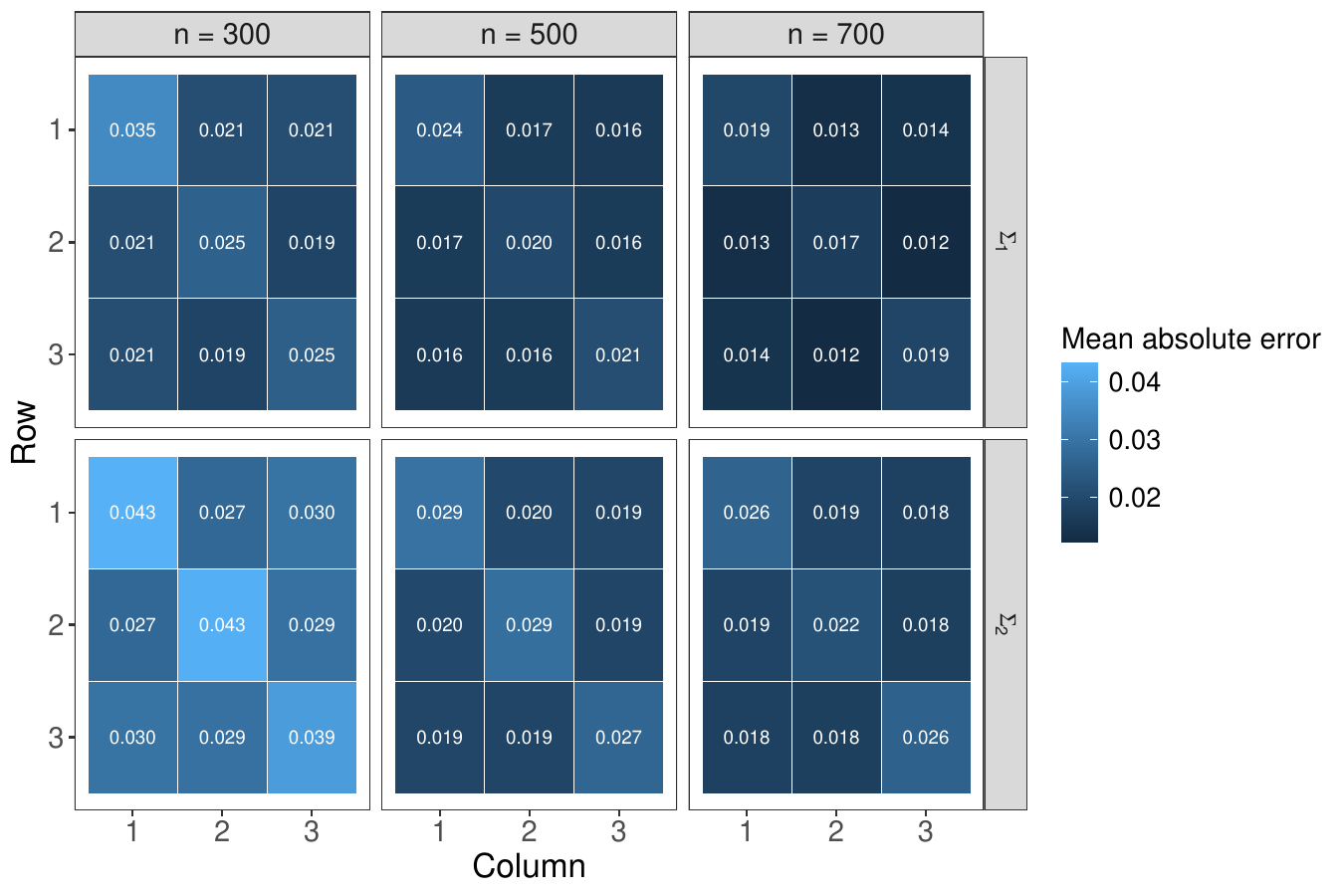}

\caption{
Entrywise covariance recovery in the \(G=2\) simulation setting.
Heatmaps show the mean absolute estimation error across 100 independent
replications, after alignment of the fitted and generating class labels,
for \textbf{(A)} the class-specific random-effect covariance matrices
\(\boldsymbol{\Omega}_1\) and \(\boldsymbol{\Omega}_2\), and
\textbf{(B)} the class-specific visit-level covariance matrices
\(\boldsymbol{\Sigma}_1\) and \(\boldsymbol{\Sigma}_2\), for
\(n\in\{300,500,700\}\). The upper and lower rows correspond to classes
1 and 2, respectively, and numerical labels give the entrywise mean
absolute errors. In panel \textbf{(A)}, grey cells represent structural
zeros imposed by the block-diagonal specification of
\(\boldsymbol{\Omega}_g\) and were not included in the error calculation.
A common color scale is used across sample sizes and classes within each
panel.
}
\label{fig:supp-g2-covariance-mae}
\end{figure}

Figure~\ref{fig:supp-g1-covariance-mae} presents the entrywise errors
for the single-class simulations. For \(\boldsymbol{\Omega}\), indices
\((1,2)\), \((3,4)\), and \((5,6)\) identify the random-intercept and
random-slope pair for outcomes 1, 2, and 3, respectively. The
indices of \(\boldsymbol{\Sigma}\) correspond directly to the three
outcomes. The displayed errors decreased across all free covariance
entries as the sample size increased from \(n=100\) to \(n=500\).
Figure~\ref{fig:supp-g2-covariance-mae} gives the corresponding results
for the two-class simulations. For both classes and both covariance
matrices, the displayed entrywise errors decreased as \(n\) increased
from 300 to 700. The errors were generally larger for class 2 than for
class 1, consistent with the blockwise relative errors reported in the
preceding subsection. Nevertheless, the errors decreased for both
diagonal and off-diagonal entries in each class, indicating improved
recovery of both the marginal variances and the dependence structure
with increasing sample size.

\subsection{Latent-class recovery with comparator methods}
\label{subsec:supp-comparator-recovery}

Because the proposed method and the comparator methods use different model
specifications, the heterogeneous simulation settings were compared in
terms of latent-class recovery rather than parameter estimation. The ARI
summaries for the proposed method are reported in
Table~\ref{tab:supp-proposed-ari}. In this subsection, we report the performance of latent-class recovery through ARI for \texttt{lcmm} and \texttt{clustGLMM}.

\subsubsection{Latent-class recovery with \texttt{lcmm}}
\label{subsubsec:supp-lcmm-ari}

The \texttt{lcmm} benchmark was fitted as described in
Section~\ref{subsec:supp-sim-comparators}. Table~\ref{tab:supp-lcmm-ari}
summarizes the resulting ARI distributions. As in the proposed-method
summary, the \(Q_1\)--\(Q_3\) column gives the interquartile interval,
whereas the IQR column gives its width, \(Q_3-Q_1\).

\begin{table}[!htbp]
\centering
\caption{ARI summaries for \texttt{lcmm} across
100 independent replications.}
\label{tab:supp-lcmm-ari}
\small
\setlength{\tabcolsep}{8pt}
\begin{tabular}{@{}ccccccc@{}}
\toprule
True \(G\) & \(n\) & Mean & Median & SD & \(Q_1\)--\(Q_3\) & IQR \\
\midrule
2 & 300 & 0.876 & 0.883 & 0.038 & 0.858--0.896 & 0.038 \\
2 & 500 & 0.890 & 0.891 & 0.027 & 0.876--0.906 & 0.030 \\
2 & 700 & 0.882 & 0.883 & 0.025 & 0.867--0.901 & 0.034 \\
3 & 500 & 0.797 & 0.798 & 0.029 & 0.774--0.818 & 0.044 \\
\bottomrule
\end{tabular}
\end{table}

For the \(G=2\) settings, \texttt{lcmm} achieved mean ARIs between
0.876 and 0.890 and median ARIs between 0.883 and 0.891. Its SD
decreased from 0.038 at \(n=300\) to 0.025 at \(n=700\), although
the mean and median ARIs were highest at \(n=500\), rather than
increasing monotonically with sample size. Comparing Tables~\ref{tab:supp-proposed-ari} and \ref{tab:supp-lcmm-ari}, the proposed method had
higher mean and median ARIs at all three sample sizes. In particular,
its median ARI exceeded that of \texttt{lcmm} by 0.026, 0.038, and
0.044 at \(n=300\), 500, and 700, respectively. The interquartile
intervals for the proposed method were also shifted upward at each
sample size, indicating consistently stronger recovery of the
generating partition.

In the more challenging \(G=3\), \(n=500\) setting, the two methods
had similar performance. The mean and median ARIs from \texttt{lcmm}
were 0.797 and 0.798, compared with 0.801 and 0.802 for the proposed
method. Both methods had an SD of 0.029. The proposed method also had
slightly higher first and third quartiles and a narrower IQR:
\(0.783\)--\(0.821\), with an IQR of 0.038, compared with
\(0.774\)--\(0.818\), with an IQR of 0.044, for \texttt{lcmm}.
Thus, compared to \texttt{lcmm}, the proposed method provided an improvement in
latent-class recovery and stability under the weaker class separation.

\subsubsection{Latent-class recovery with \texttt{clustGLMM}}
\label{subsubsec:supp-clustglmm-ari}

The \texttt{clustGLMM} benchmark was fitted as described in
Section~\ref{subsec:supp-sim-comparators}. Table~\ref{tab:supp-clustglmm-ari}
summarizes the ARI distributions from two fits for each simulation setting.
The \emph{sparse} fit used the posterior modal number \(G^+\) of occupied components,
whereas the \emph{fixed}-\(G\) fit held the number of classes at the true generating
value. As in Tables~\ref{tab:supp-proposed-ari} and
\ref{tab:supp-lcmm-ari}, the \(Q_1\)--\(Q_3\) column gives the interquartile
interval, and the IQR column gives its width, \(Q_3-Q_1\).

For every replication, the sparse fit selected the specified upper bound:
\(G^+=G_{\max}=3\) for data generated with \(G=2\), and
\(G^+=G_{\max}=4\) for data generated with \(G=3\). Consequently, the
sparse-fit rows in Table~\ref{tab:supp-clustglmm-ari} summarize the ARIs of
the selected, over-partitioned solutions, whereas the fixed-\(G\) rows
isolate latent-class recovery when the true number of classes is supplied.

\begin{table}[!htbp]
\centering
\caption{ARI summaries for \texttt{clustGLMM} under sparse and fixed-\(G\)
fits across 100 independent replications. For the sparse fits, \(G^+\) is
the posterior modal number of occupied components; for the fixed-\(G\)
fits, \(G\) was set to the true generating value.}
\label{tab:supp-clustglmm-ari}
\footnotesize
\setlength{\tabcolsep}{6pt}
\begin{tabular}{@{}cclccccc@{}}
\toprule
True \(G\) & \(n\) & Fit used for classification & Mean & Median & SD
& \(Q_1\)--\(Q_3\) & IQR \\
\midrule
2 & 300 & Sparse (\(G^+=3\)) & 0.448 & 0.444 & 0.087
& 0.400--0.504 & 0.104 \\
2 & 300 & Fixed (\(G=2\)) & 0.729 & 0.727 & 0.058
& 0.693--0.765 & 0.072 \\
\midrule
2 & 500 & Sparse (\(G^+=3\)) & 0.448 & 0.448 & 0.063
& 0.411--0.482 & 0.071 \\
2 & 500 & Fixed (\(G=2\)) & 0.733 & 0.739 & 0.040
& 0.712--0.760 & 0.048 \\
\midrule
2 & 700 & Sparse (\(G^+=3\)) & 0.446 & 0.452 & 0.068
& 0.410--0.485 & 0.075 \\
2 & 700 & Fixed (\(G=2\)) & 0.737 & 0.729 & 0.034
& 0.715--0.759 & 0.044 \\
\midrule 
3 & 500 & Sparse (\(G^+=4\)) & 0.261 & 0.264 & 0.041
& 0.234--0.290 & 0.056 \\
3 & 500 & Fixed (\(G=3\)) & 0.321 & 0.329 & 0.047
& 0.287--0.353 & 0.066 \\
\bottomrule
\end{tabular}
\end{table}

For the \(G=2\) settings, the sparse-fit partitions had mean ARIs of
0.446--0.448 and median ARIs of 0.444--0.452. Fixing the number of classes
at \(G=2\) increased the mean ARIs to 0.729--0.737 and the median ARIs to
0.727--0.739. The fixed-\(G\) results also became more concentrated as
\(n\) increased: the SD decreased from 0.058 to 0.034, and the IQR
decreased from 0.072 to 0.044. Thus, the consistent selection of an
additional class accounts for a substantial portion of the reduced ARI
obtained from the sparse fits.
Nevertheless, even with \(G\) fixed correctly, \texttt{clustGLMM} remained
below the proposed method, whose mean and median ARIs were 0.917--0.924
and 0.909--0.929, respectively, across the same three settings
(Table~\ref{tab:supp-proposed-ari}). The proposed method also gave smaller
SDs and IQRs at every sample size.

The contrast was larger in the more challenging \(G=3\), \(n=500\)
setting. The selected sparse fit had a mean ARI of 0.261 and a median ARI
of 0.264. Supplying the true \(G=3\) increased these values to 0.321 and
0.329, respectively, but they remained substantially below the mean and
median ARIs of 0.801 and 0.802 from the proposed method. Hence, fixing
\(G\) improved recovery but did not remove the performance difference
under the weaker class separation.

The difficulty of recovering the correct number of classes should be
interpreted in light of the different model-selection mechanisms and model
specifications. The sparse \texttt{clustGLMM} fit determines \(G^+\) from
the posterior number of occupied components within a model fitted with a
specified upper bound; it does not fit separate candidate models and
compare them using an information criterion. In addition, the fitted
\texttt{clustGLMM} model used constant mixing weights and did not include
the covariate-dependent gating network or the separate visit-level
multivariate Gaussian error in the data-generating model. Under these
differences, additional occupied components may accommodate
covariate-dependent class-membership patterns or residual heterogeneity
that is represented explicitly by the proposed model. The simulated over-dispersion and the overlap among
the generating trajectories, particularly for \(G=3\), may further favor
splitting a generating class into multiple occupied components.
These factors provide plausible explanations for the consistent selection
of \(G_{\max}\) in these experiments. Taken together, the sparse- and
fixed-\(G\) results indicate that both over-partitioning and differences
between the fitted and generating models contributed to the lower ARIs
from \texttt{clustGLMM}.

\section{Additional details for the COVID-19 neutrophil RNA-seq application}
\label{sec:supp-real-data}

\paragraph{Data source and cohort definition.}
We used the gene-level count matrix and corresponding sample-level metadata
and RNA-seq quality-control records from the longitudinal study of circulating
neutrophils by \citet{lasalle2022longitudinal} (Gene Expression Omnibus (GEO) accession GSE212041).
The released count matrix contained 781 RNA-seq profiles. Sample metadata and
quality-control records were matched to the count-matrix columns by the public
sample identifier, and the visit labels were required to agree between the two
metadata tables. No samples were selected according to SARS-CoV-2 status or
clinical severity.

\paragraph{Sample-level quality control and visit selection.}
We applied the six sample-level RNA-seq quality-control (QC) criteria used in the
original study. A sample was retained only if (i) the percentage of
mitochondrial reads was below 20\%; (ii) more than 10,000 genes were detected
with at least five unambiguous reads; (iii) the median exon coefficient of
variation (CV) was below 1; (iv) the median absolute deviation of the exon CV
was below 0.75; (v) the exonic rate exceeded 0.25; and (vi) the median \(3^{\prime}\)-end coverage bias, which measures
preferential read coverage near the \(3^{\prime}\) ends of transcripts,
was below 90\%. Samples with a missing value for any of these six
metrics were treated as failing quality control. In total, 698 of the 781
profiles (89.37\%) passed all six criteria, reproducing the quality-control
total reported by \citet{lasalle2022longitudinal}.

We then restricted the analysis to the scheduled hospital visits D0, D3, and
D7, coded as days 0, 3, and 7, respectively. Thus, QC-passing event-driven
draws (DE) and healthy-control profiles (H) were not included. The resulting
analysis dataset contained 662 samples from 364 subjects, with its sample and
subject composition summarized in
Table~\ref{tab:supp-covid-sample-flow}. The final cohort included both
SARS-CoV-2-positive patients and symptomatic test-negative controls, but no
healthy controls. Subjects contributed all available retained visits. An
absent scheduled visit was not imputed or interpolated.

\begin{table}[!htbp]
\centering
\caption{Sample flow for the COVID-19 neutrophil RNA-seq application.
Visit-specific counts are shown after applying all six sample-level
quality-control criteria.}
\label{tab:supp-covid-sample-flow}
\begin{tabular}{lr}
\toprule
\textbf{Stage or subgroup} & \textbf{Number} \\
\midrule
RNA-seq profiles in the released count matrix
    & 781 \\
Profiles passing all six quality-control criteria
    & 698 \\
Profiles retained at D0, D3, or D7
    & 662 \\
\quad D0
    & 348 \\
\quad D3
    & 195 \\
\quad D7
    & 119 \\
\midrule
Subjects in the final analysis cohort
    & 364 \\
\quad SARS-CoV-2 positive
    & 297 \\
\quad Symptomatic test-negative controls
    & 67 \\
\bottomrule
\end{tabular}
\end{table}

\paragraph{Count preprocessing and sequencing-depth adjustment.}
The supplied gene-level count estimates were rounded to the nearest
non-negative integer before model fitting. TMM normalization factors were
then estimated with \texttt{edgeR} from the complete count matrix for the 662
QC-passing D0, D3, and D7 samples, before restricting the response vector to
\textit{ARG1}, \textit{SERPINB2}, and \textit{ZBTB16}
\citep{robinson2010scaling,ChenEtAl2025}. The full-library sizes were retained
after gene selection. For sample \(l\) from subject \(i\), the effective
library size and known offset were
\[
L_{il}
=
\{\text{library size}\}_{il}
\times
\{\text{TMM normalization factor}\}_{il},
\qquad
o_{il}=\log(L_{il}).
\]
The same sample-specific offset \(o_{il}\) was used for each of the three
outcomes measured at that visit. The counts themselves were not divided by
\(L_{il}\) or log-transformed before fitting. Sequencing-depth and
compositional differences entered the model only through the known offset.

For model construction, visit time was scaled as
\(t_{il}=\text{day}_{il}/7\), so D0, D3, and D7 corresponded to \(0\),
\(3/7\), and \(1\), respectively. Age was represented by the five ordered
categories 20--34, 35--49, 50--64, 65--79, and 80 years or older, coded 1
through 5 and centered at its mean across the 364 retained subjects. Age was
complete in the analysis cohort and was the only clinical variable used in
model fitting. SARS-CoV-2 status, clinical severity, and comorbidity
variables were used only for post hoc description of the estimated classes.
Missing values in those descriptive variables were not imputed, that is, summaries
and tests used the available observations. The ``Unknown'' BMI category
was excluded from analyses restricted to known BMI categories.

\bibliographystyle{abbrvnat}
\bibliography{reference}

@article{subedi2025multivariate,
  title={Multivariate {P}oisson lognormal distribution for modeling counts from modern biological data: An overview},
  author={Subedi, Sanjeena and Dang, Utkarsh J},
  journal={Computational and Structural Biotechnology Journal},
  volume={27},
  pages={1255--1264},
  year={2025},
  publisher={Elsevier}
}

@article{robinson2010edger,
  title={edgeR: a {B}ioconductor package for differential expression analysis of digital gene expression data},
  author={Robinson, Mark D and McCarthy, Davis J and Smyth, Gordon K},
  journal={Bioinformatics},
  volume={26},
  number={1},
  pages={139--140},
  year={2010},
  publisher={Oxford University Press}
}

@article{love2014moderated,
  title={Moderated estimation of fold change and dispersion for {RNA-seq} data with {DESeq2} },
  author={Love, Michael I and Huber, Wolfgang and Anders, Simon},
  journal={Genome Biology},
  volume={15},
  number={12},
  pages={550},
  year={2014},
  publisher={Springer}
}

@article{inouye2017review,
  title={A review of multivariate distributions for count data derived from the {P}oisson distribution},
  author={Inouye, David I and Yang, Eunho and Allen, Genevera I and Ravikumar, Pradeep},
  journal={Wiley Interdisciplinary Reviews: Computational Statistics},
  volume={9},
  number={3},
  pages={e1398},
  year={2017},
  publisher={Wiley Online Library}
}

@article{si2014model,
  title={Model-based clustering for {RNA}-seq data},
  author={Si, Yaqing and Liu, Peng and Li, Pinghua and Brutnell, Thomas P},
  journal={Bioinformatics},
  volume={30},
  number={2},
  pages={197--205},
  year={2014},
  publisher={Oxford University Press}
}

@article{rau2015co,
  title={Co-expression analysis of high-throughput transcriptome sequencing data with {P}oisson mixture models},
  author={Rau, Andrea and Maugis-Rabusseau, Cathy and Martin-Magniette, Marie-Laure and Celeux, Gilles},
  journal={Bioinformatics},
  volume={31},
  number={9},
  pages={1420--1427},
  year={2015},
  publisher={Oxford University Press}
}

@article{papastamoulis2016estimation,
  title={On the estimation of mixtures of {P}oisson regression models with large number of components},
  author={Papastamoulis, Panagiotis and Martin-Magniette, Marie-Laure and Maugis-Rabusseau, Cathy},
  journal={Computational Statistics \& Data Analysis},
  volume={93},
  pages={97--106},
  year={2016},
  publisher={Elsevier}
}

@article{aitchison1989multivariate,
  title={The multivariate {Poisson}-log normal distribution},
  author={Aitchison, John and Ho, CH},
  journal={Biometrika},
  volume={76},
  number={4},
  pages={643--653},
  year={1989},
  publisher={Oxford University Press}
}

@article{silva2019multivariate,
  title={A multivariate {P}oisson-log normal mixture model for clustering transcriptome sequencing data},
  author={Silva, Anjali and Rothstein, Steven J and McNicholas, Paul D and Subedi, Sanjeena},
  journal={BMC Bioinformatics},
  volume={20},
  number={1},
  pages={394},
  year={2019},
  publisher={Springer}
}

@article{subedi2020family,
  title={A family of parsimonious mixtures of multivariate {P}oisson-lognormal distributions for clustering multivariate count data},
  author={Subedi, Sanjeena and Browne, Ryan P},
  journal={Stat},
  volume={9},
  number={1},
  pages={e310},
  year={2020},
  publisher={Wiley Online Library}
}

@article{silva2023finite,
  title={Finite mixtures of matrix variate {P}oisson-log normal distributions for three-way count data},
  author={Silva, Anjali and Qin, Xiaoke and Rothstein, Steven J and McNicholas, Paul D and Subedi, Sanjeena},
  journal={Bioinformatics},
  volume={39},
  number={5},
  pages={btad167},
  year={2023},
  publisher={Oxford University Press}
}

@article{payne2025finite,
  title={Finite mixtures of multivariate {P}oisson-log normal factor analyzers for clustering count data},
  author={Payne, Andrea and Silva, Anjali and Rothstein, Steven J and McNicholas, Paul D and Subedi, Sanjeena},
  journal={Statistics and Computing},
  volume={35},
  number={6},
  pages={189},
  year={2025},
  publisher={Springer}
}

@article{chiquet2021poisson,
  title={The {Poisson}-lognormal model as a versatile framework for the joint analysis of species abundances},
  author={Chiquet, Julien and Mariadassou, Mahendra and Robin, St{\'e}phane},
  journal={Frontiers in Ecology and Evolution},
  volume={9},
  pages={588292},
  year={2021},
  publisher={Frontiers Media SA}
}

@article{mcnicholas2010model,
  title={Model-based clustering of longitudinal data},
  author={McNicholas, Paul D and Murphy, T Brendan},
  journal={Canadian Journal of Statistics},
  volume={38},
  number={1},
  pages={153--168},
  year={2010},
  publisher={Wiley Online Library}
}

@article{proust2017estimation,
  title={Estimation of extended mixed models using latent classes and latent processes: the {R} package lcmm},
  author={Proust-Lima, C{\'e}cile and Philipps, Viviane and Liquet, Benoit},
  journal={Journal of Statistical Software},
  volume={78},
  pages={1--56},
  year={2017}
}

@article{daSilva2024MGLMM,
  author  = {{da Silva}, Guilherme Parreira and Laureano, Henrique Aparecido and Petterle, Ricardo Rasmussen and Ribeiro J{\'u}nior, Paulo Justiniano and Bonat, Wagner Hugo},
  title   = {Multivariate Generalized Linear Mixed Models for Count Data},
  journal = {Austrian Journal of Statistics},
  year    = {2024},
  volume  = {53},
  number  = {1},
  pages   = {44--69}
}

@article{genolini2010kml,
  title={KmL: k-means for longitudinal data},
  author={Genolini, Christophe and Falissard, Bruno},
  journal={Computational Statistics},
  volume={25},
  number={2},
  pages={317--328},
  year={2010},
  publisher={Springer}
}

@article{lasalle2022longitudinal,
  title={Longitudinal characterization of circulating neutrophils uncovers phenotypes associated with severity in hospitalized {COVID-19} patients},
  author={LaSalle, Thomas J and Gonye, Anna LK and Freeman, Samuel S and Kaplonek, Paulina and Gushterova, Irena and Kays, Kyle R and Manakongtreecheep, Kasidet and Tantivit, Jessica and Rojas-Lopez, Maricarmen and Russo, Brian C and others},
  journal={Cell Reports Medicine},
  volume={3},
  number={10},
  pages = {100779},
  year={2022},
  publisher={Elsevier}
}

@book{Bishop2006,
  author    = {Bishop, Christopher M.},
  title     = {Pattern Recognition and Machine Learning},
  series    = {Information Science and Statistics},
  publisher = {Springer},
  address   = {New York, NY},
  year      = {2006}
}

@article{WainwrightJordan2008,
  author  = {Wainwright, Martin J. and Jordan, Michael I.},
  title   = {Graphical Models, Exponential Families, and Variational Inference},
  journal = {Foundations and Trends in Machine Learning},
  year    = {2008},
  volume  = {1},
  number  = {1--2},
  pages   = {1--305}
}

@article{ArridgeEtAl2018,
  author  = {Arridge, Simon R. and Ito, Kazufumi and Jin, Bangti and Zhang, Chen},
  title   = {Variational {Gaussian} Approximation for {Poisson} Data},
  journal = {Inverse Problems},
  year    = {2018},
  volume  = {34},
  number  = {2},
  pages   = {025005}
}

@article{Schwarz1978,
  author  = {Schwarz, Gideon},
  title   = {Estimating the Dimension of a Model},
  journal = {The Annals of Statistics},
  year    = {1978},
  volume  = {6},
  number  = {2},
  pages   = {461--464}
}

@article{HubertArabie1985,
  author  = {Hubert, Lawrence and Arabie, Phipps},
  title   = {Comparing Partitions},
  journal = {Journal of Classification},
  year    = {1985},
  volume  = {2},
  number  = {1},
  pages   = {193--218}
}

@article{ChenEtAl2025,
  author  = {Chen, Yunshun and Chen, Lizhong and Lun,   Aaron T. L. and Baldoni, Pedro L. and Smyth, Gordon  K.},
  title   = {{edgeR} v4: powerful differential analysis of sequencing data with expanded functionality and improved support for small
  counts and larger datasets},
  journal = {Nucleic Acids Research},
  year    = {2025},
  volume  = {53},
  number  = {2},
  pages   = {gkaf018}
}

@article{robinson2010scaling,
  title={A scaling normalization method for differential expression analysis of {RNA}-seq data},
  author={Robinson, Mark D and Oshlack, Alicia},
  journal={Genome Biology},
  volume={11},
  number={3},
  pages={R25},
  year={2010},
  publisher={Springer}
}

@article{holm1979simple,
  title={A simple sequentially rejective multiple test procedure},
  author={Holm, Sture},
  journal={Scandinavian Journal of Statistics},
  pages={65--70},
  year={1979},
  publisher={JSTOR}
}

@article{fang2023clustering,
  title={Clustering microbiome data using mixtures of logistic normal multinomial models},
  author={Fang, Yuan and Subedi, Sanjeena},
  journal={Scientific Reports},
  volume={13},
  number={1},
  pages={14758},
  year={2023},
  publisher={Nature Publishing Group UK London}
}

@article{mclachlan2003modelling,
  title={Modelling high-dimensional data by mixtures of factor analyzers},
  author={McLachlan, Geoffrey J and Peel, David and Bean, Richard W},
  journal={Computational Statistics \& Data Analysis},
  volume={41},
  number={3-4},
  pages={379--388},
  year={2003},
  publisher={Elsevier}
}

@article{bates2015lme4,
  author  = {Bates, Douglas and M{\"a}chler, Martin and Bolker, Ben
             and Walker, Steve},
  title   = {Fitting Linear Mixed-Effects Models Using {lme4}},
  journal = {Journal of Statistical Software},
  year    = {2015},
  volume  = {67},
  number  = {1},
  pages   = {1--48}
}

@article{vavra2024clusterwise,
  author  = {V{\'a}vra, Jan and Kom{\'a}rek, Arno{\v{s}}t
             and Gr{\"u}n, Bettina and Malsiner-Walli, Gertraud},
  title   = {Clusterwise Multivariate Regression of Mixed-Type Panel Data},
  journal = {Statistics and Computing},
  year    = {2024},
  volume  = {34},
  number  = {1},
  pages   = {1--20}
}

@book{ripley2002modern,
  title     = {Modern Applied Statistics with S},
  author    = {Venables, William N. and Ripley, Brian D.},
  year      = {2002},
  edition = {4\textsuperscript{th}},
  publisher = {Springer},
  address   = {New York}
}

@manual{nnet,
  title  = {\texttt{nnet}: Feed-Forward Neural Networks and
            Multinomial Log-Linear Models},
  author = {Ripley, Brian and Venables, William},
  year   = {2025},
  note   = {R package version 7.3-20},
  doi    = {10.32614/CRAN.package.nnet},
  url    = {https://CRAN.R-project.org/package=nnet}
}

\end{document}